\pdfoutput=1
\documentclass[cernpreprint,texlive=2011,txfonts,UKenglish,PAPER]{latex/atlasdoc}

\usepackage{latex/atlaspackage}
\usepackage{latex/atlasbiblatex}

\usepackage{latex/atlasphysics}

\usepackage{siunitx}

\graphicspath{{logos/}{figures/}}

\usepackage{type3seesaw-defs}
\usepackage{amsmath}

\hypersetup{pdftitle={Search for heavy lepton resonances decaying to a $Z$ boson and a lepton in $pp$ collisions at $\sqrt{s}=8\TeV$ with the ATLAS detector},pdfauthor={The ATLAS Collaboration}}
\newcommand{\mytitle}{Search for heavy lepton resonances decaying to a $Z$ boson and a lepton in $pp$ collisions at $\sqrt{s}=8\TeV$ with the ATLAS detector}

\AtlasTitle{\mytitle}

\author{The ATLAS Collaboration}

\AtlasRefCode{EXOT-2014-08}

\PreprintIdNumber{CERN-PH-EP-2015-080}

\AtlasJournal{\textcolor{black}{JHEP}}

\newcommand{\myabstract}{A search for heavy leptons decaying to a $Z$ boson and an electron or a muon is presented. The search is based on $pp$ collision data taken at $\sqrt{s}=8$~\TeV\ by the ATLAS experiment at the CERN Large Hadron Collider, corresponding to an integrated luminosity of 20.3~fb$^{-1}$. Three high-transverse-momentum electrons or muons are selected, with two of them required to be consistent with originating from a $Z$ boson decay.  No significant excess above Standard Model \textcolor{black}{background} predictions is observed, and 95\% confidence level limits on the production cross section of high-mass trilepton resonances are derived.  The results are interpreted in the context of vector-like lepton and type-III seesaw models. For the vector-like lepton model, most heavy lepton mass values in the range $114$--$176 \GeV$ are excluded. For the type-III seesaw model, most mass values in the range $100$--$468 \GeV$ are excluded.}

\AtlasAbstract{\myabstract}

\begin{document}

\tableofcontents

\section{Introduction}
\label{sec:intro}

Searches for new particles often utilize decays to electrons or muons, which allow for full four-momentum reconstruction with a mass resolution better than that achievable using hadronic or semileptonic decay modes. Dilepton resonance searches led to the discovery of the $J/\psi$~\cite{jpsi1,jpsi2}, the ${\varUpsilon}$~\cite{upsilon}, and the $Z$ boson~\cite{zua1, zua2}, and they have been used at the CERN Large Hadron Collider (LHC) to place strong constraints on a variety of new particles such as additional gauge bosons~\cite{CMS_dilepton,zprime_8TeV}. Searches for low-mass trilepton resonances have been used to constrain lepton flavour violation in muon and $\tau$ lepton decays~\cite{Bellgardt:1987du,taulll}.

High-mass trilepton resonances are motivated by several extensions of the Standard Model (SM).
Vector-like leptons (VLL) are invoked to explain the mass hierarchy between the different lepton generations~\cite{Falkowski:2013jya}.
They also arise in composite Higgs models~\cite{Fujikawa:1994we,KerenZur:2012fr} and models of warped extra dimensions~\cite{Redi:2013pga,Contino:2006nn}.
Such leptons have masses much larger than those of the SM leptons, and are defined as colourless, spin-1/2 fermions whose left- and right-handed chiral components have the same transformation properties under the weak-isospin $\mathrm{SU}(2)$ gauge group.
Another set of models predicting trilepton resonances is based on the type-III seesaw mechanism~\cite{t3seesaw}, which explains the origin of small neutrino masses through the introduction of heavy $\mathrm{SU(2)}$ triplets with zero hypercharge.

This article presents a search for high-mass trilepton resonances with the ATLAS detector, using a data sample corresponding to $20.3~\ifb$ of integrated luminosity collected in $pp$ collisions at $\sqrt{s}=8\TeV$ at the LHC. This search uses data events with at least three charged leptons (electrons or muons), two of which are consistent with originating from a $Z$-boson decay. Several signal regions are defined to be sensitive to the pair-production of heavy leptons that decay to SM leptons and $W$, $Z$, or $H$ bosons. The backgrounds, dominated by SM diboson production, are estimated using Monte Carlo (MC) simulation and control regions in data, and the predictions are validated in dedicated data samples.

The results of the search are interpreted in the context of vector-like lepton~\cite{hgg} and type-III seesaw~\cite{Biggio:2011ja} scenarios in which the new heavy leptons decay through mixing with electrons or muons ($\ell$) induced by off-diagonal Yukawa couplings.
In the type-III seesaw model, the masses of the three heavy leptons are assumed to be identical.
 Feynman diagrams of the production and decay of the heavy leptons in both models are shown in figure~\ref{fig:heavy-lepton-feynman-diagrams}.
 The heavy leptons are produced in pairs through Drell--Yan processes, with cross sections of roughly 34~fb and 844~fb for the VLL and type-III seesaw models, respectively, assuming heavy lepton masses of 200~\GeV.
  The difference in the production cross section is due to the different gauge couplings of the models, as well as the additional neutral fermion in the type-III seesaw model.
  The heavy leptons decay via the mixing terms into an SM lepton and a $W$, $Z$, or $H$ boson.
 The charged states $\lpm$ exist in both models, and have decay modes to $W\nu$, $Z\ell$, and $H\ell$; the neutral state $\lzero$ is only present in the type-III seesaw model, and decays to $W\ell$, $Z\nu$, and $H\nu$.
The charged lepton branching fractions approach ${\cal B}(\lpm\to W^{\pm}\nu)=50\%$, ${\cal B}(\lpm\to Z\ell^{\pm})=25\%$ and ${\cal B}(\lpm\to H\ell^{\pm})=25\%$ for $m_{\lpm}\gg m_H$, in accordance with the Goldstone boson equivalence theorem~\cite{Chanowitz1985379}; at lower masses, the branching fractions to $H$ and $Z$ decrease as they become kinematically disfavoured.
 For the neutral lepton, the branching fractions to $W\ell$, $Z\nu$, and $H\nu$ are identical to those of the charged leptons to $W\nu$, $Z\ell$, and $H\ell$, respectively.

\begin{figure}[htbp]
  \centering
  \subfloat[ ] {
    \resizebox{0.45\textwidth}{!}{\includegraphics{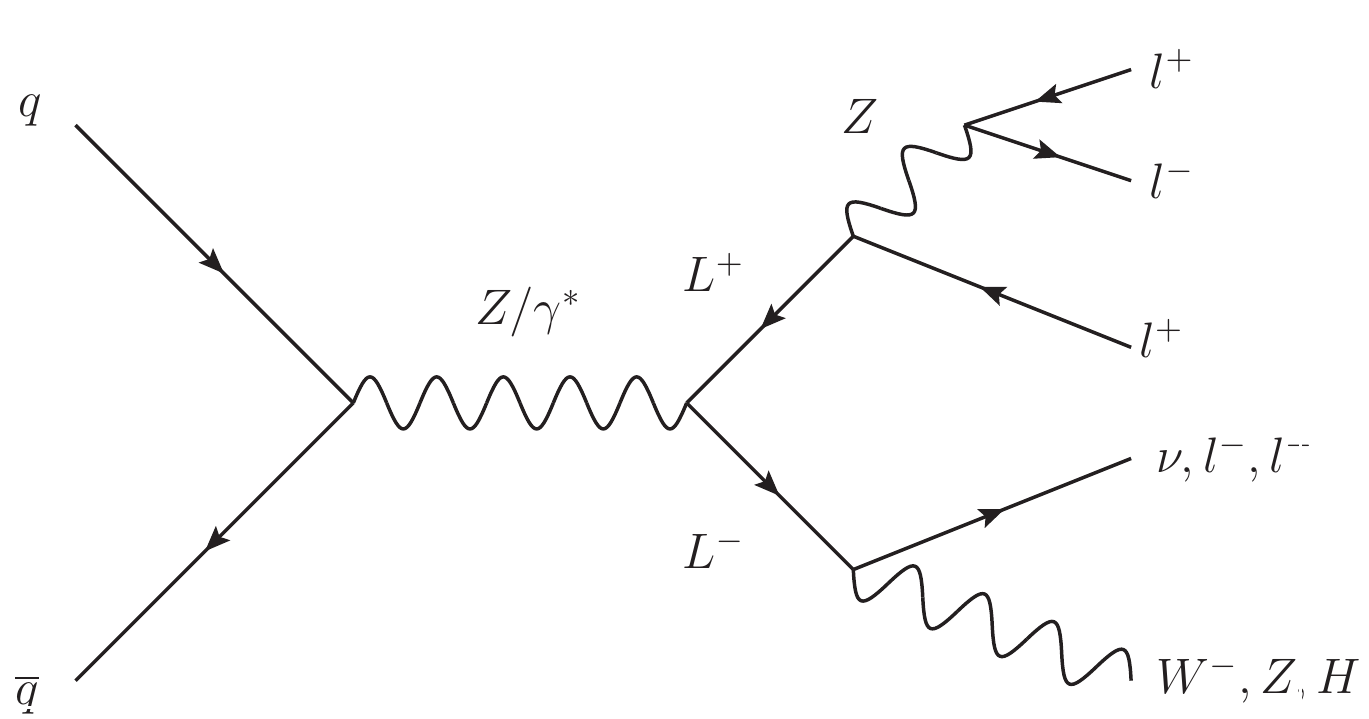}}
  }
  \subfloat[ ] {
    \resizebox{0.45\textwidth}{!}{\includegraphics{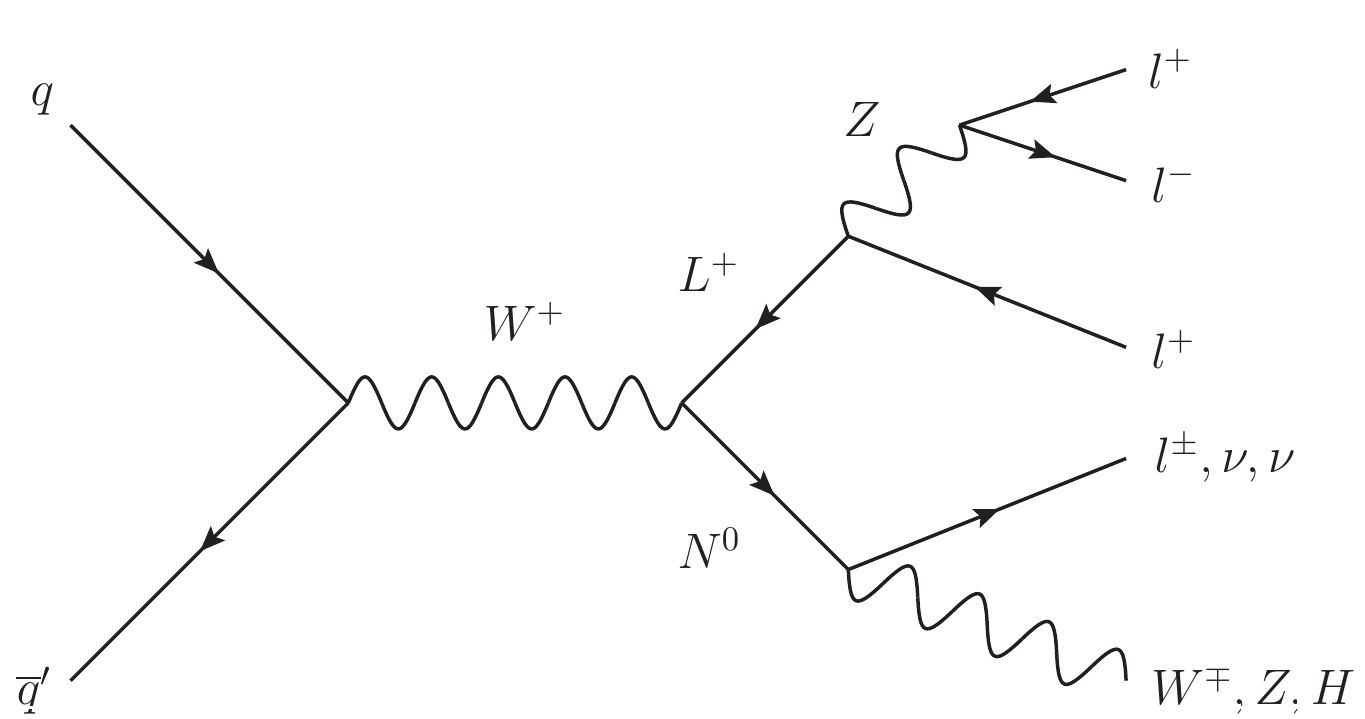}}
  }
  \caption{Feynman diagrams for the production and decay of new heavy leptons ($\lpm$, $\lzero$) to final states resulting in a trilepton resonance. Diagram (a) shows the pair production of two charged heavy leptons, and (b) shows the associated production of a charged and a neutral heavy lepton.}
  \label{fig:heavy-lepton-feynman-diagrams}
\end{figure}

Searches for heavy leptons were previously performed at LEP, excluding vector-like leptons with masses below $m_{\lpm}=101.2~\GeV$ using the $L^{\pm}\rightarrow W^{\pm}\nu$ decay mode~\cite{Achard:507268}. A search for type-III seesaw heavy leptons was performed by CMS in $pp$ collision data at $\sqrt{s}=7~\TeV$, using non-resonant trilepton signatures to exclude seesaw fermions with masses below $m_{\lpm}=180$--$200 \GeV$, depending on the branching fractions assumed~\cite{Chatrchyan2012348}. 

\section{The ATLAS detector}
\label{sec:detector}
The ATLAS detector~\cite{atlas} is a multi-purpose detector covering nearly the full solid angle~\footnote{ATLAS uses a right-handed coordinate system with its origin at the nominal interaction point (IP) in the centre of the detector and the $z$-axis along the beam pipe. The $x$-axis points from the IP to the centre of the LHC ring, and the $y$-axis points upward. Cylindrical coordinates $(r,\phi)$ are used in the transverse plane, $\phi$ being the azimuthal angle around the beam pipe. The pseudorapidity is defined in terms of the polar angle $\theta$ as $\eta=-\ln\tan(\theta/2)$.} around the $pp$ interaction region. The beam pipe is surrounded by the inner detector (ID), consisting of silicon pixel and microstrip detectors and a transition radiation tracker. The ID is enclosed in a superconducting solenoid providing a 2~T axial magnetic field, and performs charged particle tracking for $|\eta|<2.5$.

The calorimeter system surrounds the solenoid, and consists of electromagnetic and hadronic components. The electromagnetic calorimeter is a lead/liquid argon (LAr) sampling calorimeter, and comprises a barrel ($|\eta|<1.475$) and two endcaps ($1.375<|\eta|<3.2)$. In the range $|\eta|<2.5$, the detector is finely segmented in $\eta$ to provide good spatial resolution. The hadronic calorimeter (HCAL) uses steel/scintillator tiles in the barrel ($|\eta|<1.7$) and copper/LAr in the endcaps ($1.5<|\eta|<3.2$). In the forward region ($3.1<|\eta|<4.9$), electromagnetic and hadronic calorimetry is performed using copper/LAr and tungsten/LAr technology.

The muon spectrometer (MS) features high-precision tracking chambers interleaved with dedicated trigger chambers located in a toroidal magnetic field. The magnetic field is generated by a system of three large superconducting air-core toroid magnets, with a bending integral of about $2.5$~T$\cdot$m in the barrel and up to $6$~T$\cdot$m in the endcaps. The precision tracking is provided by monitored drift tubes ($|\eta|<2.7$), complemented by cathode strip chambers in the forward region ($2<|\eta|<2.7$). Triggering is performed by resistive plate chambers in the barrel ($|\eta|<1.05$) and thin gap chambers in the endcaps ($1.05<|\eta|<2.4$).

Events are recorded using a three-level trigger system. The first level, implemented in hardware, reduces the event rate to less than $75~\mbox{kHz}$ using a subset of the detector information. The second and third levels are implemented in software, and reduce the event rate to less than $400~\mbox{Hz}$ using the full detector information.

\section{Object reconstruction and event selection}
\label{sec:selection}

The data were collected during 2012 using triggers requiring either an electron or a muon with transverse momentum relative to the beam axis, \pt, greater than 24~\GeV. The triggered electron or muon must also satisfy loose isolation requirements. These triggers are supplemented by triggers without isolation requirements, but with higher $\pt$ thresholds of $60$ ($36$)~\GeV{} for electrons (muons). {Only data taken while the ID, calorimeters, and MS were functioning normally are considered.} Events are required to have a reconstructed primary vertex having at least three associated tracks with $\pt>400 \MeV$, consistent with the beamspot envelope. If more than one such vertex is found, the vertex with the largest $\sum p_{\mathrm{T}}^{2}$ of its associated tracks is chosen as the hard-scatter primary vertex. 

Electron candidates are selected as energy clusters within a small window of size $\Delta\eta\times\Delta\phi=0.075\times0.125$ in the electromagnetic calorimeter matched to a track in the ID. They are required to fulfil tight identification criteria~\cite{elecperf}, have $|\eta_{\mathrm{cluster}}|<2.47$, and not be in the transition region between the barrel and the endcap calorimeter ($1.37<|\eta_{\mathrm{cluster}}|<1.52$), where $\eta_{\mathrm{cluster}}$ is the pseudorapidity of the barycentre of the energy cluster. Muon candidates are selected as tracks reconstructed in the MS matched to tracks in the ID~\cite{muonperf} and are required to satisfy $|\eta|<2.5$. The muon momentum is determined from combining the information from the two tracks. Muons and electrons are required to have transverse momenta greater than 15~\GeV{} and to be isolated from tracks and calorimeter energy deposits using the criteria described in ref.~\cite{multileptons8TeV}. To ensure that the lepton track is consistent with originating from the primary event vertex, the ID track is required to satisfy $|d_0/\sigma_{d_0}|<3$ and $|z_0\sin\theta|<0.5~\mbox{mm}$, where $d_0$ and $z_0$ are the transverse and longitudinal impact parameters of the track with respect to the primary vertex, respectively, and $\sigma_{d_0}$ is the uncertainty on the transverse impact parameter. In order to ensure constant trigger efficiency as a function of lepton $\pt$, at least one electron or muon must have $\pt>26$~\GeV{} and a separation $\Delta R=\sqrt{(\Delta \eta)^2 + (\Delta \phi)^2}$ less than $0.2$ from the triggered electron or muon. The trigger efficiency is evaluated to be larger than 95\% when all offline selection criteria are applied. 

Jets are reconstructed from topological clusters built from energy deposits in calorimeter cells using the anti-$k_t$ jet algorithm~\cite{antikt} with a radius parameter of $R=0.4$. 
The measured jet energy is calibrated using $\pt$- and $\eta$-dependent corrections for instrumental effects (e.g. passive material and non-compensating response of the calorimeters) derived from MC simulations and in situ techniques applied to data, and is corrected for additional $pp$ interactions per bunch crossing (pileup)~\cite{jetcorr,JER}.

{After energy calibration,} jets are required to have $\pt>30$~\GeV{} and $|\eta|<4.5$. In order to suppress jets from pileup interactions, if a jet has $\pt<50 \GeV$ and $|\eta|<2.5$, then at least 50\% of the scalar sum of the $\pt$ of all tracks associated with the jet is required to come from tracks associated with the primary event vertex. Jets are also required to pass jet-quality selections to reject jets reconstructed from non-collision signals, such as beam-related background, cosmic rays or detector noise.

Since leptons and jet candidates can be reconstructed as multiple objects, the overlap between the various objects is resolved by applying the following procedure. If two electrons are separated by $\Delta R < 0.1$, the electron with the lower $\pt$ is removed. If an electron and a jet are separated by $\Delta R < 0.2$, the jet is removed. If an electron and a jet satisfy $0.2 < \Delta R < 0.4$, and the jet's transverse momentum also satisfies $p_{\mathrm{T},\mathrm{jet}} > 30~\mbox{GeV} + 0.05 p_{\mathrm{T},\mathrm{e}}$, the electron is removed. If a muon and an electron satisfy $\Delta R < 0.1$, the electron is removed.
If a muon and a jet are separated by $\Delta R < 0.1$, the jet is removed if its transverse momentum satisfies $p_{\mathrm{T},\mathrm{jet}}<0.5 p_{\mathrm{T},\mu}$ if $p_{\mathrm{T},\mu}<200~\mbox{GeV}$, or $p_{\mathrm{T},\mathrm{jet}} < 100~\mbox{GeV}$ if $p_{\mathrm{T},\mu}>200~\mbox{GeV}$.
Finally, if a muon and a jet not removed by the previous requirement are separated by $\Delta R < 0.3$, the muon is removed.

{The missing transverse momentum, $\textit{{\textbf{p}}}_{\mathrm{T}}^{\mathrm{miss}}$, and its magnitude, $\MET$, are calculated from the vector sum of the transverse momenta of all calibrated electrons, muons, $\tau$ leptons, jets, and all topological calorimeter clusters of energy not associated with other objects with $|\eta |<4.9$.}

Events are required to have at least three leptons (electrons or muons) passing the selection requirements above. 
At least one pair of leptons with the same flavour and opposite electric charge must have an invariant mass within 10~\GeV{} of the $Z$ boson mass, $m_Z$~\cite{pdg}. Events with four leptons consistent with the decay of two $Z$ bosons, also within 10~\GeV{} of $m_Z$, are vetoed. For the remaining events with four leptons, the lepton closest in $\Delta R$ to the $Z$ boson candidate, referred to here as the off-$Z$ lepton, is used to form the trilepton mass.  For the range of heavy lepton masses considered in this analysis, the $Z$ boson and the off-$Z$ lepton tend to be collimated; hence, to improve the signal to background ratio, events where the $Z$ candidate and the off-$Z$ lepton are separated by $\Delta R>3$ are vetoed.

For simulated events with an $\lpm$ decaying to three leptons with $\pt>15$~\GeV{} and
$|\eta|<2.5$, of which two originate from a $Z$ boson and have an invariant mass within $10 \GeV$ of $m_Z$, the efficiency of this event selection for the $Z+e$ ($Z+\mu$) decay channel ranges from 20\% (36\%) at $m_{\lpm}=100$~\GeV{} to 35\% (38\%) at $m_{\lpm}=400$~\GeV.  The determination of the efficiency is discussed in section~\ref{sec:results}.

Since the heavy leptons are produced in pairs, in addition to the identified $\lpm \to Z+\ell$ decay, signal events contain either a second $\lmp$ or an $\lzero$, which decays to a $W$, $Z$, or $H$ boson and a charged or neutral lepton. A large fraction of events therefore contain a fourth lepton and/or a hadronically
decaying boson. The sensitivity of the analysis is significantly improved by
separating the events selected above into the following three exclusive categories:
\begin{itemize}
\item ${\fourl}$: at least four leptons are required using the same identification
  criteria as described above.
\item ${\threeljj}$: exactly three leptons are required, along with two jets with an invariant mass satisfying {$m_W-20 \GeV < m_{\mathrm{jj}} < 150 \GeV$, where $m_W$ is the $W$} boson mass~\cite{pdg}.
\item ${\threelo}$: the event does not fulfil the criteria of either the
  $\fourl$ or the $\threeljj$ categories.
\end{itemize}

Subdividing the $\fourl$ category based on the presence of a dijet does not significantly improve the sensitivity due to the small number of expected events with both a fourth lepton and a hadronically decaying boson. Finally, events are separated into two channels based on whether the off-$Z$ lepton is an electron or a muon.  This classification results in six independent signal regions. 

The search is performed by looking for a narrowly peaked excess of events in the distributions of the mass difference defined by ${\deltam \equiv m_{3\ell} - m_{\ell^{\pm}\ell^{\mp}}}$, where the invariant mass of the two leptons associated with the $Z$-boson decay is subtracted from the trilepton invariant mass. This reduces the impact of the lepton momentum resolution,
and thus enhances the narrow resonance structure of the signal.  The resulting reconstructed width in $\deltam$ is $5.9 \GeV$ ($15.5 \GeV)$ for a mass hypothesis of $m_{\lpm}=120 \GeV$ ($400 \GeV$), while the corresponding width in the trilepton invariant mass is $7.3 \GeV$ ($18 \GeV$), for final states where the off-$Z$ lepton is an electron.  For final states where the off-$Z$ lepton is a muon, the corresponding width in $\deltam$ is $5.1 \GeV$ ($31.5 \GeV$) for a mass hypothesis of $m_{\lpm} = 120 \GeV$ ($400 \GeV$), while the corresponding width in the trilepton invariant mass is $6.7 \GeV$ ($33.5 \GeV$).  The intrinsic width of the resonance is a few \MeV{} at $m_{\lpm}=120 \GeV$, rising to $0.5 \GeV$ at $m_{\lpm}=400 \GeV$.

\section{Monte Carlo simulation} 
\label{sec:simulation}

The analysis uses MC samples of VLL and type-III seesaw events generated with \madgraph~4.5.2 and 5.2.2.1~\cite{madgraph}, respectively, using the CTEQ6L1~\cite{ct6l1} parton distribution functions (PDF) and the AU2 underlying event tune~\cite{AU2}. Showering is performed with \pythia\ 8~\cite{Sjostrand:2007gs}. Decays of the heavy leptons in the VLL model are performed using \bridge~\cite{bridge}, while decays in the type-III seesaw samples are performed by \madgraph. For the type-III seesaw model, the charged and neutral heavy leptons are generated with identical masses. Vector-like lepton samples are generated for eleven mass hypotheses for $100 \GeV \leq m_{\lpm}\leq400 \GeV$, while the type-III seesaw samples are generated for ten mass hypotheses for $100 \GeV \leq m_{\lpm,\lzero} \leq 500 \GeV$. The cross sections for both samples are calculated at leading order (LO) in QCD.

The main backgrounds originate from SM diboson production, in particular $WZ$ and $ZZ$ production. Contributions from $WZ$ ($ZZ$) are modelled using the \sherpa~\cite{sherpa} MC generator version 1.4.3 (1.4.5), using the internal showering algorithm~\cite{Hoeche:2009rj,Gleisberg:2008fv,Schumann:2007mg},  with the CT10~\cite{ct10} PDF set and normalized to the next-to-leading-order (NLO) prediction from \vbfnlo-2.6.2~\cite{vbfnlo}.  The generation includes up to three additional parton emissions in the matrix element. Samples of simulated events based on the NLO generator \powheg~\cite{powheg} are used to derive systematic uncertainties on the shapes of distributions predicted by \sherpa. The diboson samples are showered with \pythia~8, and use the CT10 PDF set and AU2 underlying event tune.

Drell--Yan production in association with a photon that converts in the detector, denoted $Z+\gamma$, is modelled using \sherpa~1.4.1, also using the CT10 PDF set and including up to three additional parton emissions in the matrix element. Production of top-quark pairs in association with a $W$ or $Z$ boson ($\ttV$) and triboson production ($VVV^{(*)}$) are modelled using \madgraph~5.1.3.33, with \pythia~6.426 for the parton shower and hadronization, AUET2B underlying event tune~\cite{AUET2B}, and the CTEQ6L1 PDF set. The $\ttV$ processes are normalized to the corresponding NLO cross sections~\cite{Campbell:2012dh,Lazopoulos:2008de}, while the $Z+\gamma$ and $VVV^{(*)}$ processes are normalized to their LO cross sections from the respective generator.

For all samples, the response of the ATLAS detector is modelled using the \geant\ toolkit~\cite{geant,atlassimulation}. Additional $pp$ interactions in the same or nearby bunch crossings are included in the simulation by overlaying minimum-bias interactions modelled with \pythia~6.425 onto the hard-scatter event. The simulated events are reweighted to reproduce the distribution of the average number of $pp$ interactions per crossing observed in data. The generator, parton shower, PDF set, underlying event tune, and accuracy of theoretical cross section for the primary MC samples used are summarized in table~\ref{table:mc}.

\begin{table}
  \centering
  \caption{Summary of the primary signal and background MC samples used in this analysis. The generator, parton shower and hadronization, PDF, underlying event tune, and the order of the cross-section calculation are shown for each sample.}
  \begin{tabular}{|c|c|c|c|c|c|}
    \hline
    Process & Generator & Parton shower and hadr. & PDF set & UE tune & Cross section \\
    \hline
    VLL & \madgraph\ 4.5.2 & \pythia\ 8 & CTEQ6L1 & AU2	&	LO 	\\
    Seesaw & \madgraph\ 5.2.2.1 & \pythia\ 8 & CTEQ6L1 & AU2	&	LO 	\\
    $WZ$ & \sherpa 1.4.3 & \sherpa & CT10 & \sherpa 	&	NLO\\
    $ZZ$ & \sherpa 1.4.5 & \sherpa & CT10 & \sherpa 	&	NLO\\
    $\ttbar+W/Z$ & \madgraph\ 5.1.3.33  & \pythia\ 6.426   & {CTEQ6L1} & AUET2B 	&	NLO\\
    $VVV^{(*)}$ & \madgraph\ 5.1.3.33 & \pythia\ 6.426 & {CTEQ6L1}  & AUET2B 	& 	LO	\\
    $Z+\gamma$ & \sherpa  & \sherpa & CT10 & \sherpa 	&	LO 	\\
    \hline
  \end{tabular}
  \label{table:mc}
\end{table}

\section{Background estimation} 
\label{sec:backgrounds}

Standard Model processes containing three or more lepton candidates can be classified into two categories. The first category consists of events with three prompt leptons produced in the decays of electroweak gauge bosons, which are estimated using the simulated samples described above. The second consists of events where at least one reconstructed lepton arises from a misidentified jet, hadron decay, or photon conversion, and is referred to as reducible background. For muons, reducible backgrounds arise from semileptonic $b$- or $c$-hadron decays and from in-flight decays of pions or kaons. Reducible electron backgrounds can arise from semi-leptonic $b$- or $c$-hadron decays, photon conversions and misidentified hadrons or jets. Drell--Yan production of a lepton pair with an associated photon that converts in the detector and is reconstructed as an isolated lepton ($Z+\gamma$) is estimated using simulation. The remainder of the reducible background is estimated by scaling control samples in data, following a method similar to that described in ref.~\cite{ATLAS:2014kca}. The control samples consist of events with one or more leptons that do not satisfy the nominal selection criteria, but instead satisfy a set of relaxed criteria, defined separately for each lepton flavour. The events are weighted with scale factors computed for each such lepton, defined as the ratio of misidentified or non-prompt lepton candidates that satisfy the nominal criteria to those which only fulfil the relaxed criteria. For electrons, the identification requirement is changed from tight to loose~\cite{elecperf}.  For muons, the requirements on the lepton isolation and on $|d_{0}/\sigma_{d_{0}}|$ are relaxed.
The scale factors are measured as a function of the candidate's $\pt$ and $\eta$ in samples of data that are enriched in non-prompt and misidentified lepton candidates. 
The contamination from prompt leptons in the background-enriched samples is removed using simulation.

The background estimates are validated in four validation regions. The high-$\Delta R$ region consists of events where the $Z$ boson candidate and the off-$Z$ lepton are separated by $\Delta R>3$. The background composition in this region is similar to that in the signal regions. The off-$Z$ region contains events with exactly three leptons, where no opposite-sign same-flavour pair of leptons is reconstructed with an invariant mass within 20~\GeV{} of $m_Z$.  This region is designed to test the $Z+\gamma$ background estimate.  The $ZZ$ region consists of events with two reconstructed $Z$ boson candidates. These three regions are completely orthogonal to the signal selection. The fourth region is designed to test the modelling of SM $WZ$ production.  Events in this region must have exactly three leptons, zero jets, $40 < \met < 100 \GeV$, and $40 < \mtw < 90 \GeV$, where $\mtw = \sqrt{2p_{\mathrm{T}}^\ell\met(1-\cos(\Delta\phi))}$ is the transverse mass and $\Delta\phi$ is the azimuthal angle between the missing transverse momentum and the off-$Z$ lepton with momentum $p_{\mathrm{T}}^{\ell}$.  This region is not completely orthogonal to the signal regions, but signal processes are expected to account for less than $3\%$ of the expected event yield for type-III seesaw leptons with $m_{\lpm}>160 \GeV$.  

The expected and observed numbers of events are given in table~\ref{table:cr-summary} for all validation regions, separately for the $Z+e$ and the $Z+\mu$ channels. The largest difference is seen in the off-$Z$ region in the $Z+\mu$ flavour channel, where there is a deficit in the data corresponding to $2.3$ standard deviations ($\sigma$). The region is dominated by contributions from $ZZ$, where only three leptons pass the selection requirements and no same-flavour, opposite-sign lepton pair is reconstructed with invariant mass within $20 \GeV$ of $m_Z$. 
 In the other seven regions, agreement better than $1.5\sigma$ is observed. Figure~\ref{fig:cr} shows the $\deltam$ distributions for the high-$\Delta R$ and $ZZ$ validation regions in the $Z+e$ and the $Z+\mu$ flavour channels. 

\begin{table}[htbp]
  \centering
  \caption{Summary of the number of events observed and predicted for each validation region. The uncertainty on the background prediction is the total systematic uncertainty. The difference between the observed and predicted number of events divided by the combined statistical and systematic uncertainty on the prediction is also shown.}
  \begin{tabular}{|c|c|c| r@{\ $\pm$\ }l |c|}
    \hline
    Channel     & Validation Region             & Data          & \multicolumn{2}{c|}{Background Prediction} & $\frac{\mbox{Data}-\mbox{Bkgd}}{\sigma_{\mathrm{bkgd}}}$ \\
    \hline
    $Z+e$	&	High-$\Delta R$	        &	$239$	&~~~~~~~~~~$239$&$14$ 	&    \ \ 0.0	\\
    $Z+e$	&	Off-$Z$		        &	$360$	&$349$&$44$ 	&	+0.2	\\
    $Z+e$	&	$ZZ$		        &	$39$	&$37$&$2$       &	+0.3	\\
    $Z+e$	&	$WZ$			&	$140$	&$133$&$10$ 	&	+0.4	\\
    \hline
    $Z+\mu$	&	High-$\Delta R$	        &	$302$	&$301$&$12$ 	&	+0.1	\\
    $Z+\mu$	&	Off-$Z$		        &	$163$	&$200$&$8$ 	&	$-$2.3	\\
    $Z+\mu$	&	$ZZ$		        &	$74$	&$63$&$3$ 	&	+1.2	\\
    $Z+\mu$	&	$WZ$			&	$222$	&$193$&$14$ 	&	+1.5	\\
    \hline
  \end{tabular}
  \label{table:cr-summary}
\end{table}

\begin{figure}[htbp]
  \centering
  \subfloat[ ] {
    \resizebox{0.5\textwidth}{!}{\includegraphics{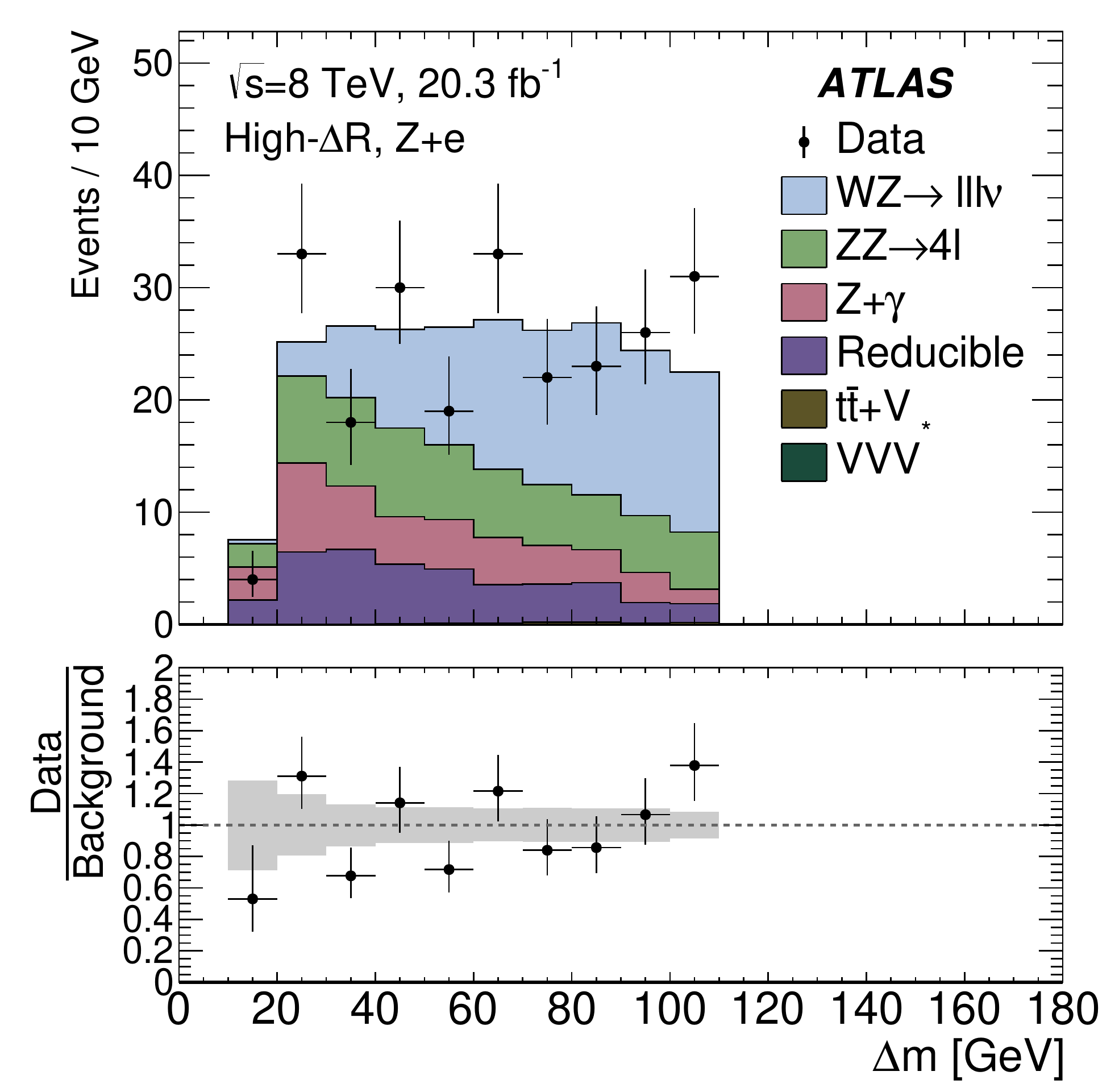}}
  }
  \subfloat[ ] {
    \resizebox{0.5\textwidth}{!}{\includegraphics{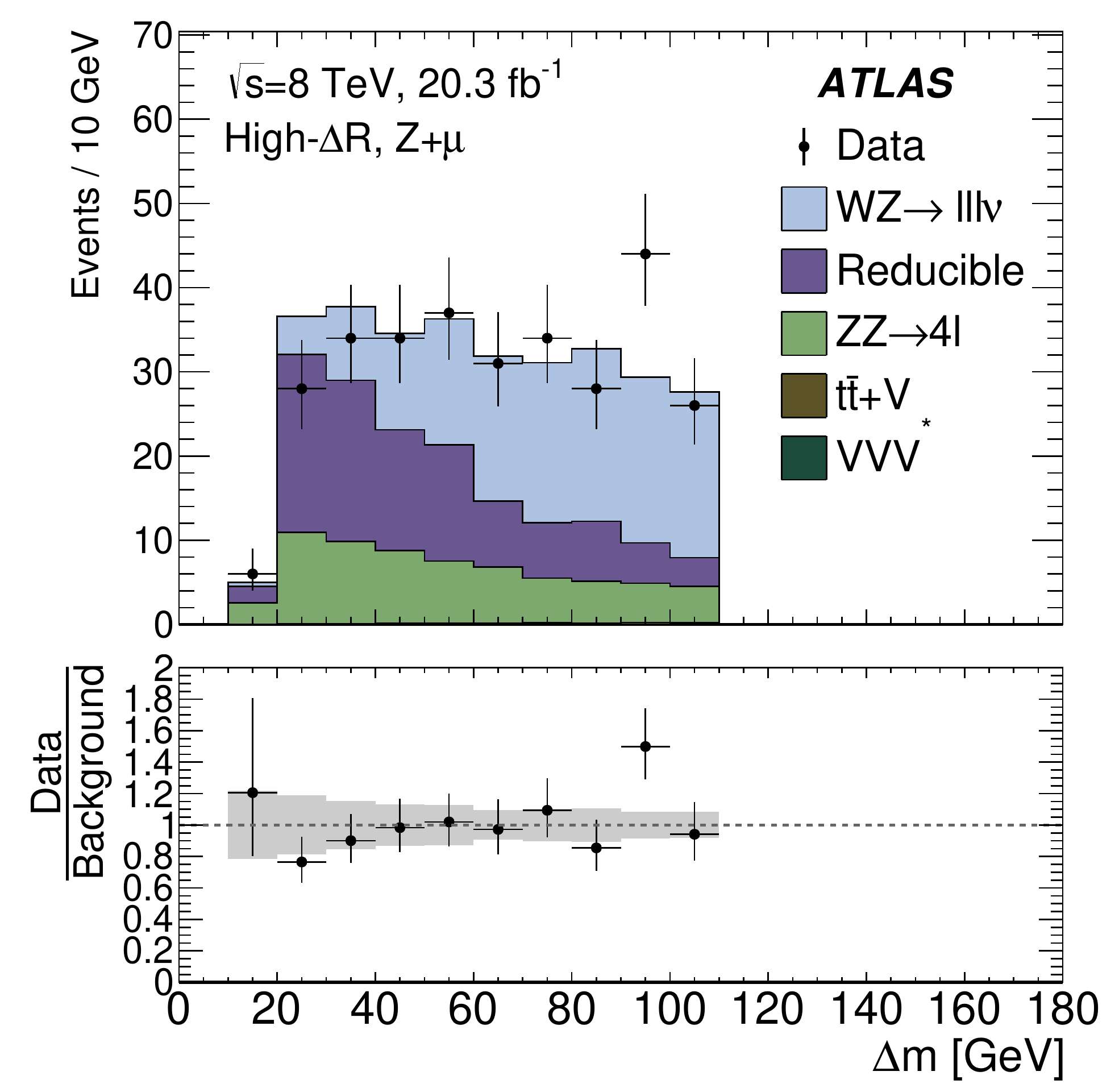}}
  }\\
  \subfloat[ ] {
    \resizebox{0.5\textwidth}{!}{\includegraphics{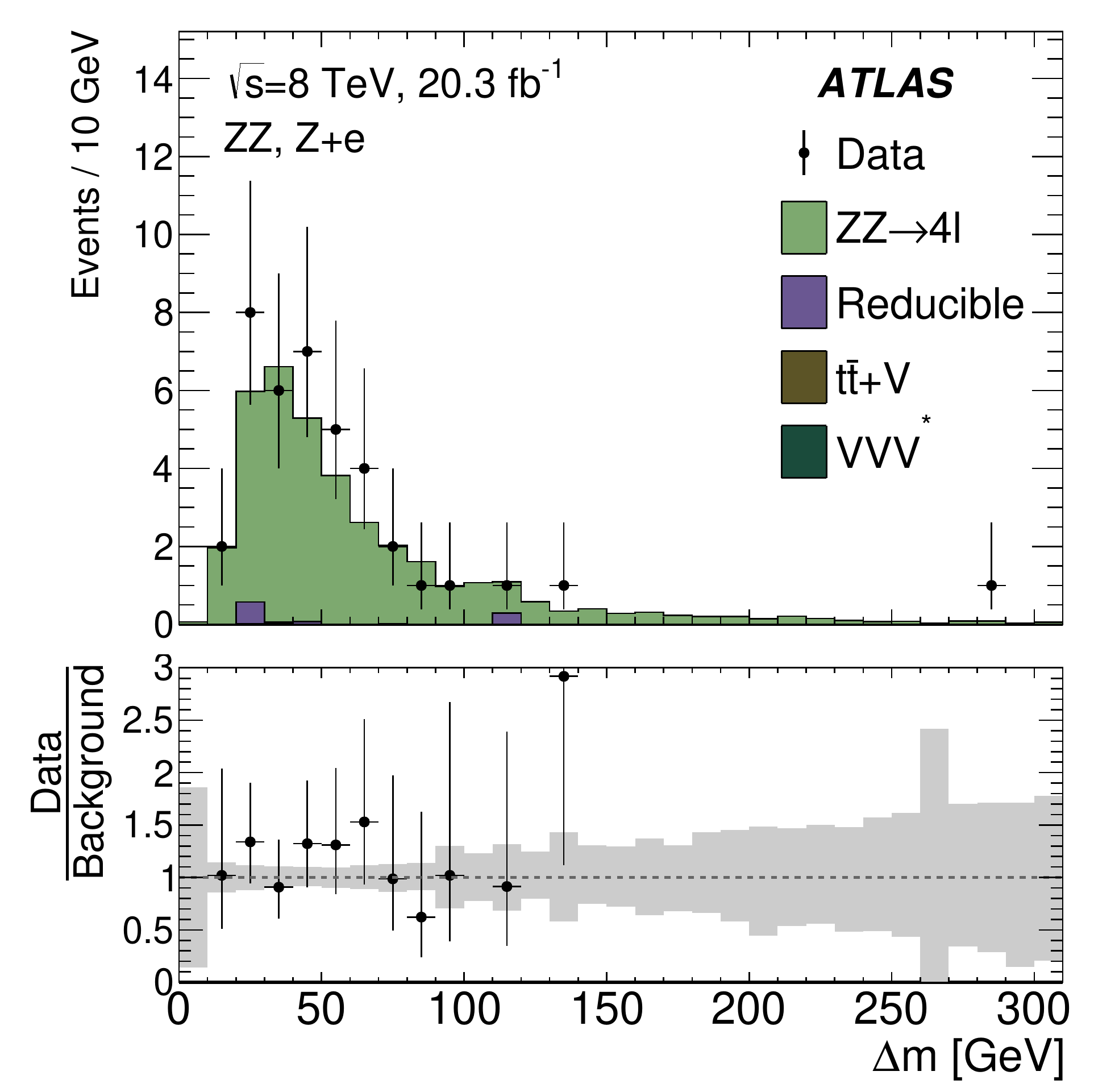}}
  }
  \subfloat[ ] {
    \resizebox{0.5\textwidth}{!}{\includegraphics{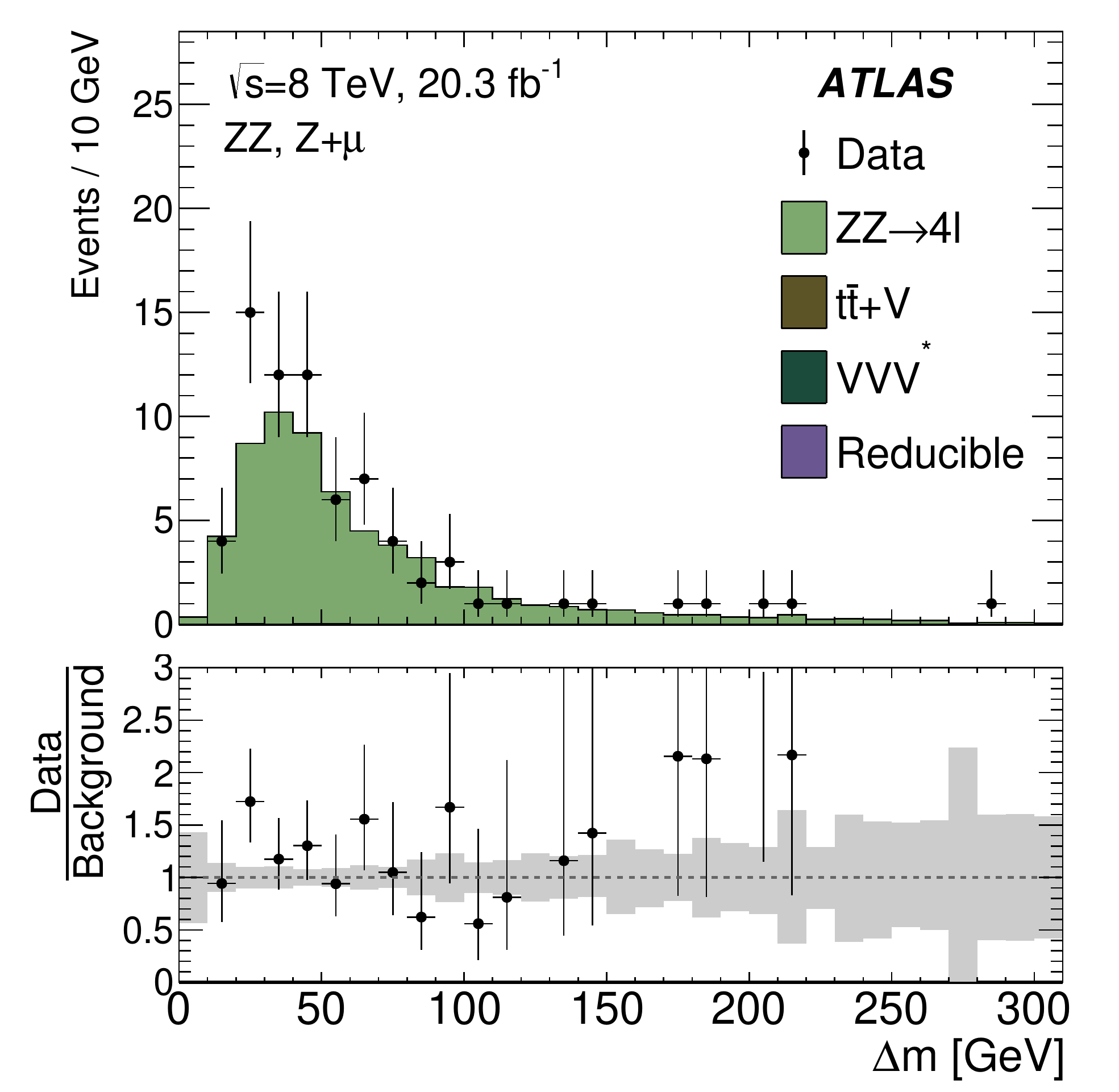}}
  }\\
  \caption{The $\deltam = m_{3\ell}-m_{\ell^+\ell^-}$ distributions for the high-$\Delta R$ validation region (a) and (b) and the $ZZ$ validation region (c) and
    (d). The figures (a) and (c) show the distributions for the $Z+e$ final states, while (b) and (d) show the $Z+\mu$ final states. The error bars on the data points represent statistical uncertainties, and the shaded band represents the systematic uncertainty on the background prediction.}
  \label{fig:cr}
\end{figure}

\section{Systematic uncertainties}
\label{sec:systematics}
Systematic uncertainties are assigned to the signal and background predictions derived from simulation to account for possible modelling inaccuracies. Sources of uncertainty affecting all simulated signal and background processes are the cross sections of SM processes, trigger efficiencies, lepton energy scales and resolutions (LES/LER), jet energy scale and resolution (JES/JER), lepton reconstruction and selection efficiencies, MC statistical uncertainties, and luminosity. The cross-section uncertainties, evaluated for the SM background samples, include renormalization and factorization scale and PDF uncertainties.  The scale uncertainties are determined by varying the renormalization and factorization scales up and down by factors of two. PDF uncertainties are obtained using the PDF4LHC working group recommendations~\cite{Botje:2011sn}.  Scale and PDF uncertainties are added in quadrature to obtain the total uncertainty on the inclusive cross section. For the dominant $WZ$ and $ZZ$ backgrounds, the resulting theoretical uncertainty on the NLO predictions from \vbfnlo{} are $7.6\%$ and $4.3\%$, respectively. A further uncertainty is assigned to the $WZ$ and $ZZ$ backgrounds to account for potential generator-level mismodelling of the shape of the $\deltam$ spectrum. The uncertainty is the difference between the predictions from \sherpa{} and \powheg, symmetrized around the value from \sherpa. 

For the $Z+\gamma$ backgrounds, an uncertainty of 30\% is assigned to the modelling of prompt photons converting in the inner detector{, based on comparisons of conversion processes in $Z\rightarrow ee$ events between data and simulation}. The reducible backgrounds are assigned an uncertainty related to the data-driven scaling procedure described in section~\ref{sec:backgrounds}, primarily due to the extrapolation of the scale factors from the measurement sample to the signal regions and to the correction for the presence of prompt leptons in the background-enriched samples. The uncertainties on the electron factors range from $24\%$ to $30\%$ as a function of $\pt$, and the uncertainties on the muon factors range from $25\%$ to $50\%$. 

The uncertainties on the lepton reconstruction and selection efficiencies, energy scales, and energy resolutions~\cite{elecperf2012,muonperf,elecES} affect all simulated backgrounds, with combined uncertainties of $1\%$ to $2\%$ on the normalizations. The jet energy scale and resolution uncertainties~\cite{JES,JER} only significantly affect the $\threeljj$ signal regions, with a total uncertainty of $3\%$. Statistical uncertainties due to the finite number of events in the MC samples range from $1\%$ to $5\%$. The luminosity uncertainty is $2.8\%$, and is derived using the same methodology as that described in ref.~\cite{luminosity}. In total, the systematic uncertainties on the background predictions in each signal region range from $6\%$ to $9\%$. 

The largest sources of uncertainty affecting the signal predictions are the lepton reconstruction and selection efficiencies, the luminosity, and, for the $\threeljj$ category, the jet energy scale and resolution. The total uncertainties on the signal normalizations range from $3\%$ to $7\%$ depending on the signal region and $m_{\lpm}$.

\section{Signal and background model}
\label{sec:fit-procedure}

The numbers of signal and background candidate events in data are determined from an unbinned maximum-likelihood fit of a combination of signal and background models to the $\deltam$ distributions in each signal region. The details of the fit procedure and the models are described below.

The signal and background processes are modelled with probability density functions (p.d.f.s). The parameters of the p.d.f.s are determined from fits to the background estimates described in section~\ref{sec:backgrounds}. The fit to data using the combined signal and background model is performed simultaneously on the three categories for each of the two flavour channels. In each signal region, the normalization of the dominant background ($ZZ$ or $WZ$) is a free parameter in the fit. The normalizations of all other backgrounds are constrained to fluctuate according to Gaussian probability distributions with mean and width values equal to the estimates and the total uncertainties before fitting. The uncertainties on the p.d.f. parameters are incorporated as Gaussian-distributed nuisance parameters.

The VLL and type-III seesaw signal models are parameterized separately as the sum of a Voigtian function (the convolution of a Breit--Wigner and a Gaussian function) for the trilepton resonance peak and a Landau distribution for the combinatorial part of the signal, where the three reconstructed leptons do not originate from the same $\lpm$ decay. The signal parameterization at a certain heavy lepton mass $m_{\lpm}$ is given as a function of $\deltam$ by the following expression:
\begin{equation}
  \mathcal{S}(m_{\lpm}) = f_{\mathrm{V}} F_{\mathrm{V}}(\deltam; \Gamma_{\mathrm{V}}, m_{\mathrm{V}}, \sigma_{\mathrm{V}} )
  + (1-f_{\mathrm{V}})F_{\mathrm{L}}(\deltam; m_{\mathrm{L}}, \sigma_{\mathrm{L}}),
\end{equation}
where $f_{\mathrm{V}}$ denotes the fraction of events in the resonance peak (Voigtian function), $\Gamma_{\mathrm{V}}$, $m_{\mathrm{V}}$ and $\sigma_{\mathrm{V}}$ are the width, mean, and Gaussian smearing terms of the Voigtian function $F_{\mathrm{V}}$, and $m_{\mathrm{L}}$ and $\sigma_{\mathrm{L}}$ are the parameters of the Landau distribution $F_{\mathrm{L}}$. The six parameters are determined at each simulated mass point by fitting $\mathcal{S}(m_{\lpm})$ to the simulated $\deltam$ distributions, separately for the two flavours.  For mass points $m_{\lpm}$ that lie between those assumed in the MC samples, the parameters of the signal templates are obtained by linearly interpolating the fitted values determined at the nearest simulated mass points. {The fraction of events in the Voigtian part of the signal, $f_V$, is $\sim60 \% (\sim 70\%)$ at $120\GeV$, decreasing to $\sim 58 \% ( \sim 55\%)$ at $400\GeV$ for the type-III seesaw signal (VLL).} The uncertainties on the fit parameters of the signal p.d.f. are incorporated as Gaussian-distributed nuisance parameters in the combined fits to data. 

The combined background model consists of five different p.d.f.s, corresponding to $WZ$, $ZZ$, $Z+\gamma$, reducible, and the sum of the $t\bar{t}+V$ and triboson backgrounds. The leading $WZ$ and $ZZ$ backgrounds are both modelled with a modified Bukin function~\cite{Bukin}, a three-parameter function designed to model peaks with asymmetric tails using the convolution of a Gaussian and an exponential function. To mitigate the impact of MC statistical uncertainties, the parameterizations for the $4\ell$ and $\threelo$ categories are determined from the combination of all three categories; for the $\threeljj$ category, a separate parameterization is used to account for possible kinematic effects from the two additional jets. The uncertainty on the shape of the $\deltam$ distribution predicted by the generator is taken into account with a Gaussian-distributed nuisance parameter multiplying a template given by the difference between the p.d.f.s determined from \sherpa\ and \powheg. Finally, in the $\threeljj$ and $\threelo$ categories, the ratio of normalizations of the $WZ$ and $ZZ$ backgrounds is fixed to the prediction from MC simulation, due to the inability of the fit to resolve the similar shapes of the $\deltam$ distributions. In the $4\ell$ category, the contribution from $WZ$ events is negligible, so only the $ZZ$ background normalization is left as a free parameter.

The most important remaining backgrounds are due to reducible processes and $Z+\gamma$ production. The reducible background is parameterized with a Landau distribution, determined from the data-driven estimate described above. Due to the low number of events in the reducible  background estimates, the corresponding shape is obtained by fitting the combined distribution of all six signal regions.

The $Z(\ell\ell)+\gamma$ background contributes significantly only to the $Z+e$, $\threelo$ category. This background is modelled with the sum of a Landau distribution and a Gaussian function.  The normalizations of the reducible and $Z+\gamma$ backgrounds are constrained to the expected values, each with a Gaussian-distributed uncertainty of 30\%.

Finally, the background contributions from triboson and $t\bar{t}+W/Z$ production are modelled together using a Landau distribution, due to their similar shape and small contribution to the total background estimate. The normalization in the combined fit is also constrained to the expected value from simulation, with a 30\% Gaussian-distributed uncertainty.

\section{Results}\label{sec:results}

The total number of events observed in each signal region is shown in table~\ref{table:SR-event-counts}, along with the estimated backgrounds 
before and after fitting the total background model to the data.
The corresponding $\deltam$ distributions for the pre-fit background estimates and the data are shown in figure~\ref{fig:SR-DeltaM}. The signals expected for the VLL model with $m_{\lpm}=140 \GeV$ and the type-III seesaw model with $m_{\lpm}=300 \GeV$ are superimposed on the background as illustrative examples. The data agree with the background expectation in all cases, and no clear peak indicating resonant trilepton production is seen in any of the signal regions. 

\begin{figure}[htbp]
  \centering
  \subfloat[] {
    \includegraphics[width=0.4\columnwidth]{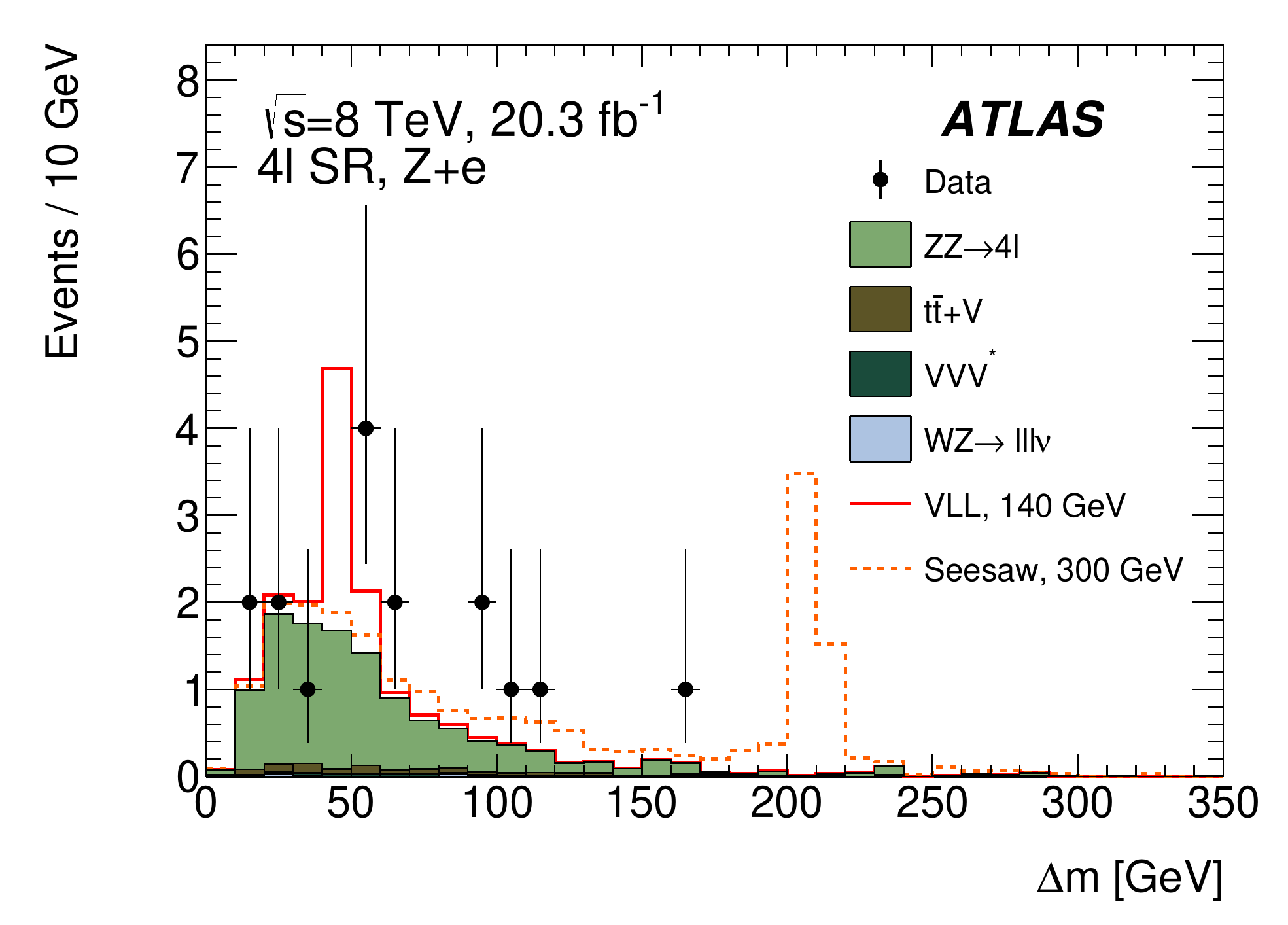}
  }
  \subfloat[] {
    \includegraphics[width=0.4\columnwidth]{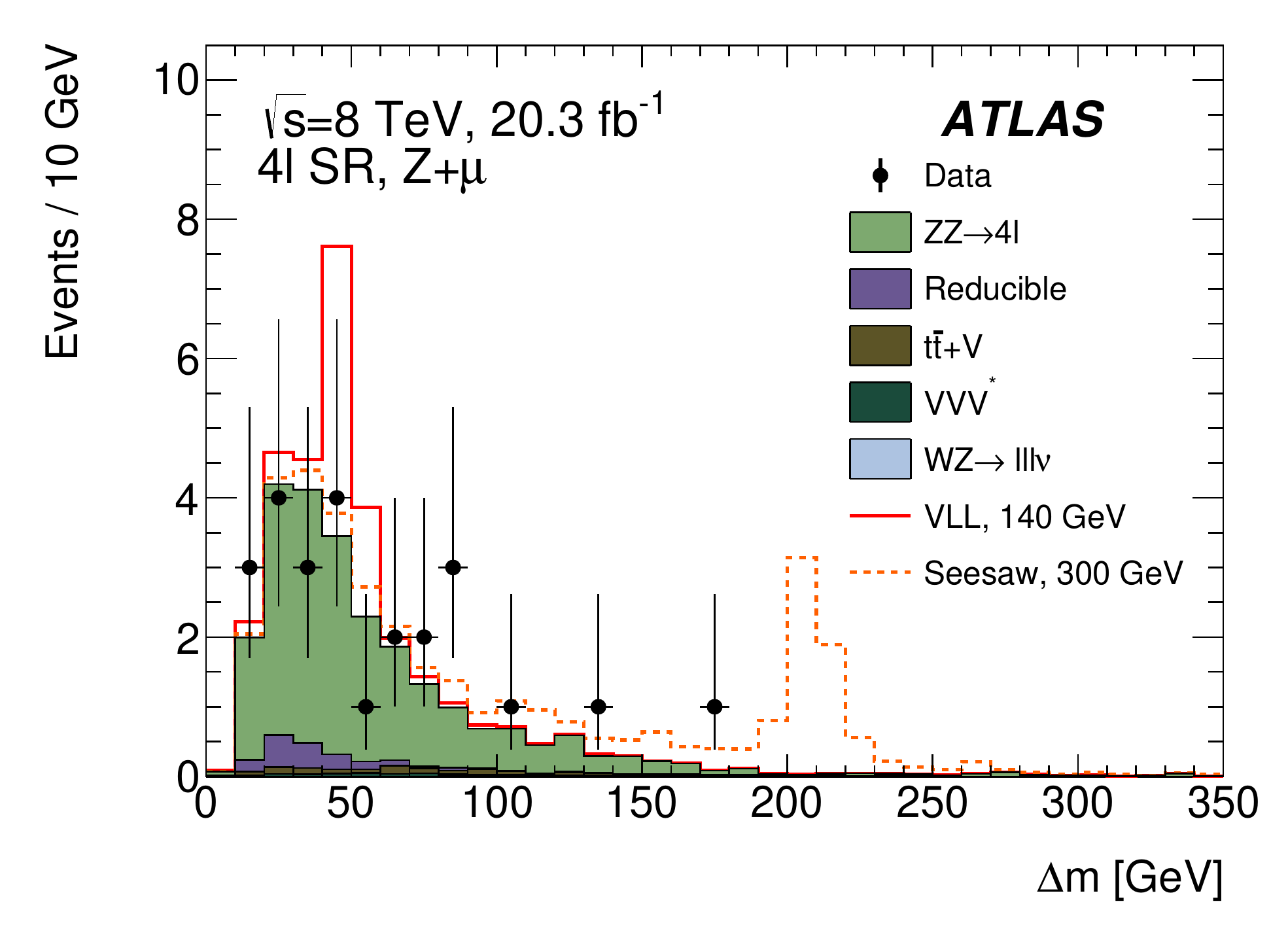}
  } \\
  \subfloat[] {
    \includegraphics[width=0.4\columnwidth]{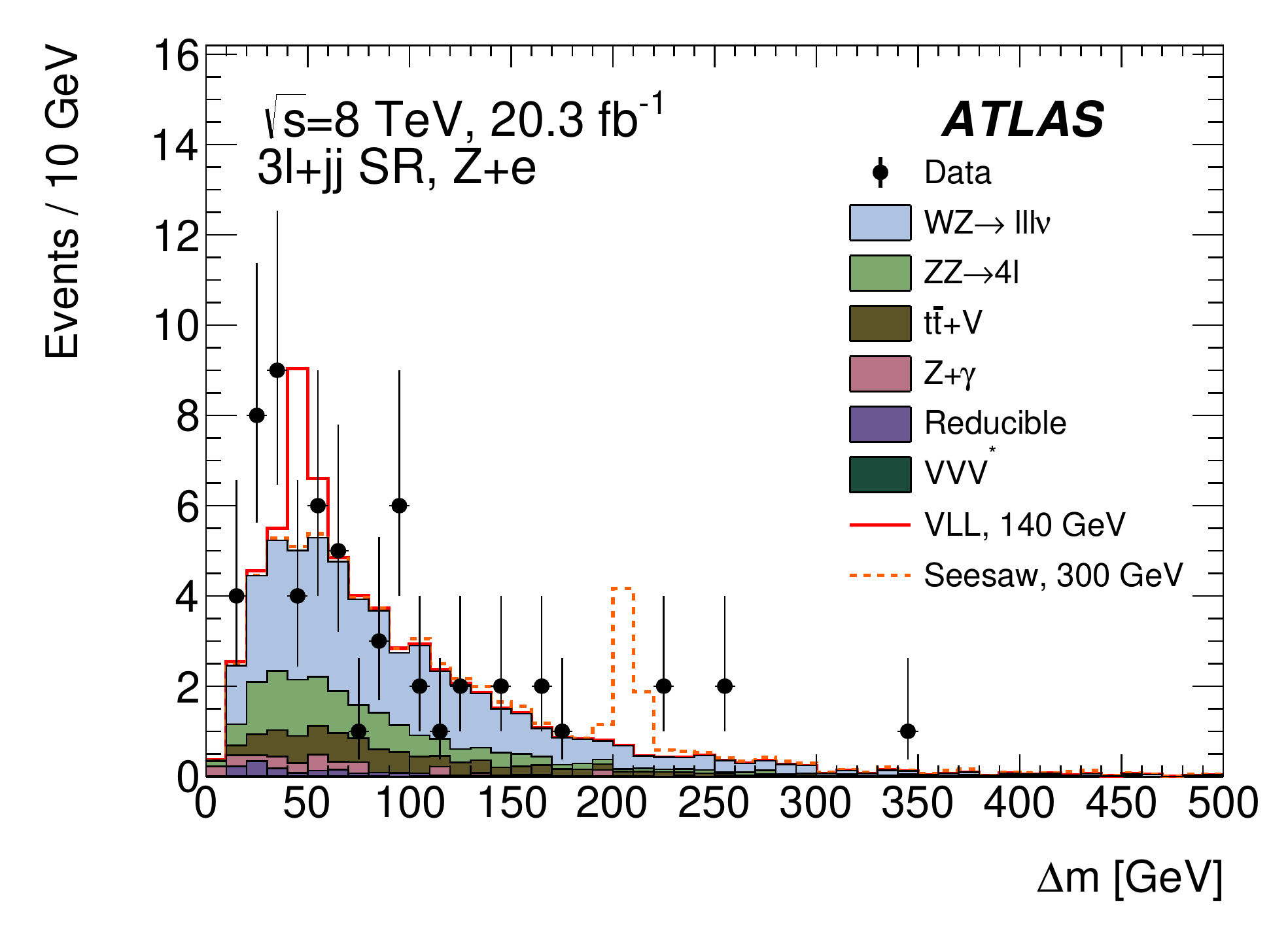}
  }
  \subfloat[] {
    \includegraphics[width=0.4\columnwidth]{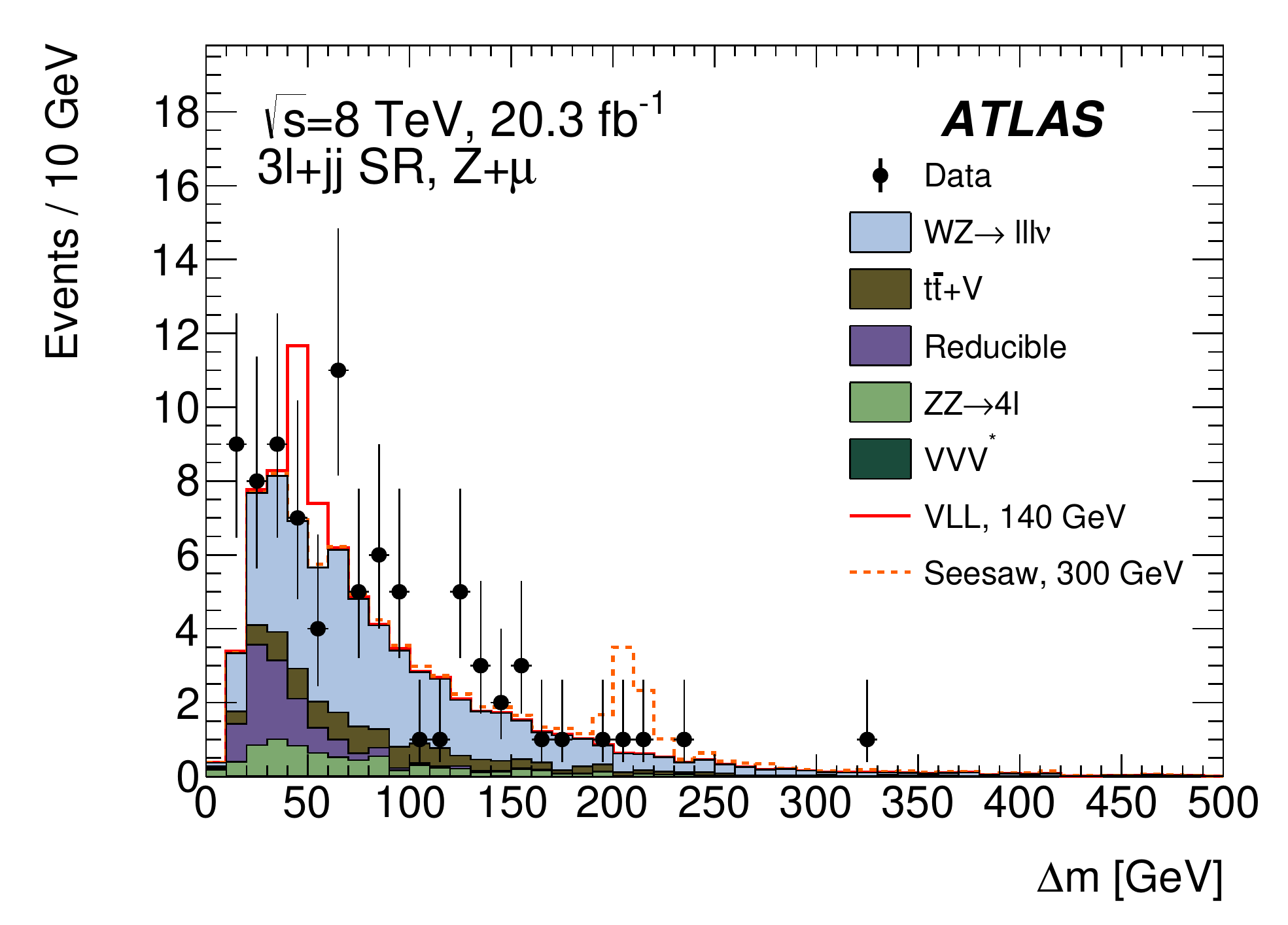}
  } \\
  \subfloat[] {
    \includegraphics[width=0.4\columnwidth]{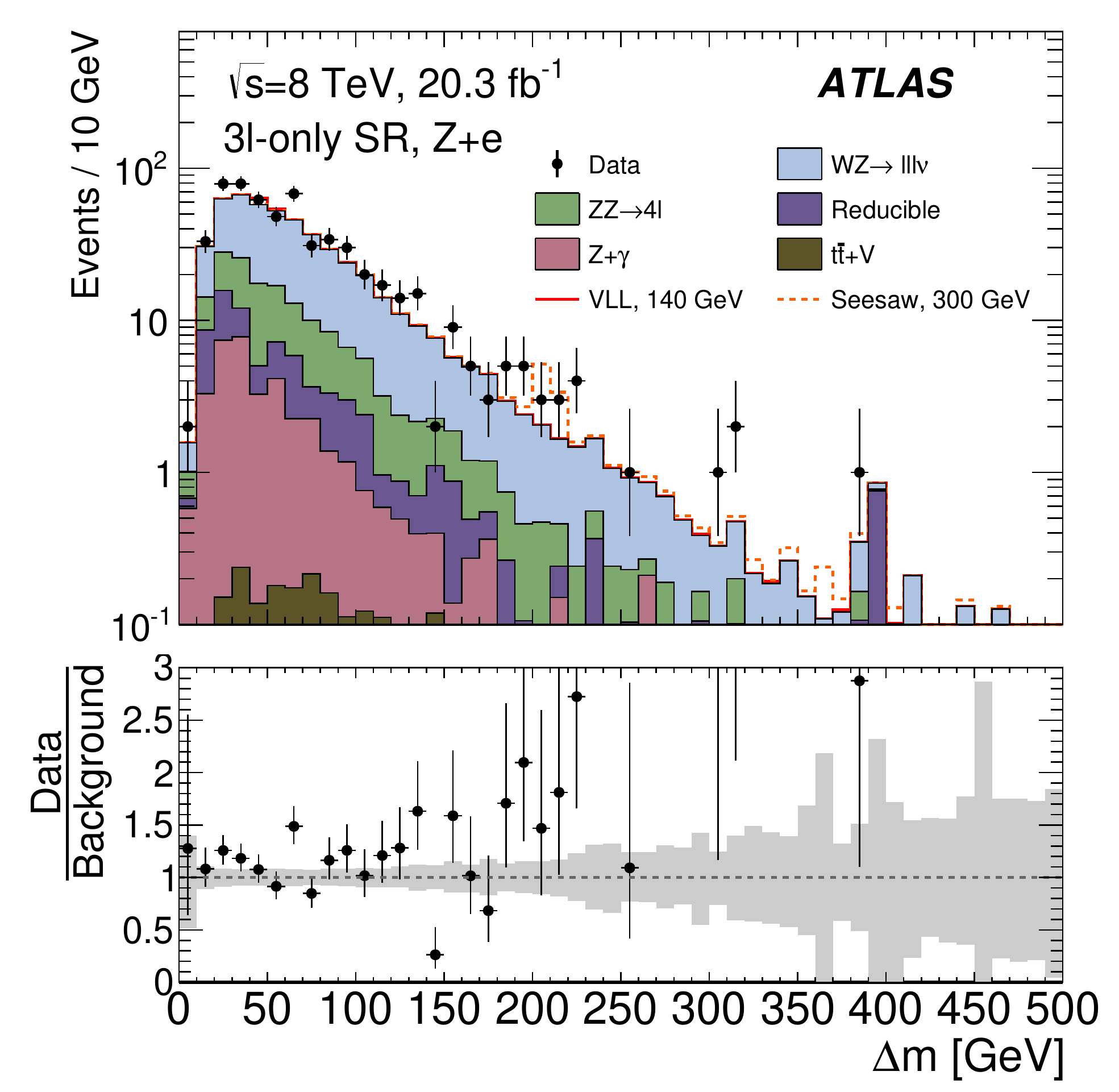}
  }
  \subfloat[] {
    \includegraphics[width=0.4\columnwidth]{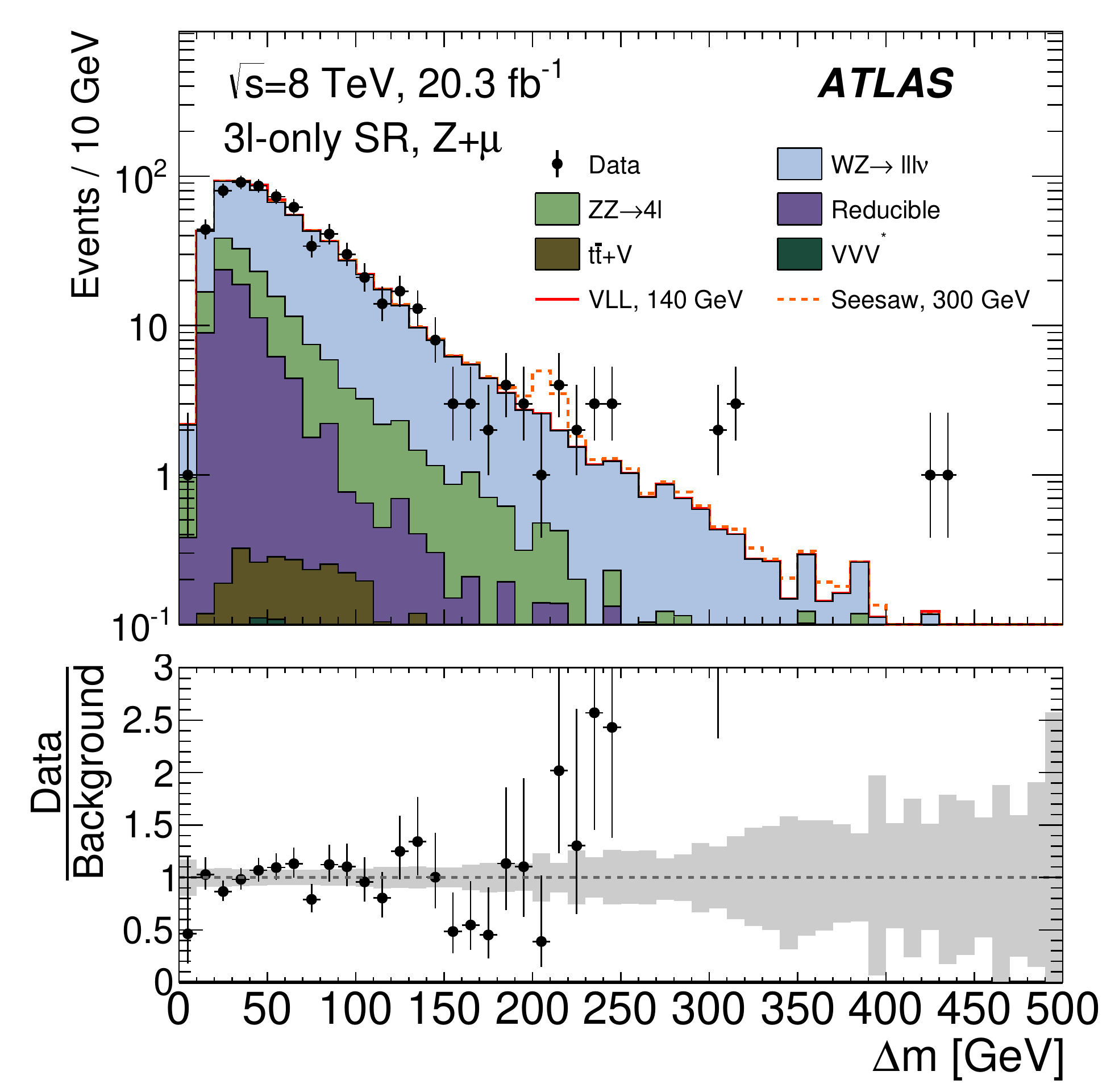}
  }
   \caption{The $\deltam = m_{3\ell}-m_{\ell^+\ell^-}$ distributions for the $\fourl$ (top), $\threeljj$ (middle), and $\threelo$ (bottom) categories, divided into the $Z+e$ (left) and $Z+\mu$ (right) flavour channels. The observed data are shown as black points, while the pre-fit background expectations are shown in the coloured histograms. Also shown are examples for signal contributions for a 140~GeV $\lpm$ in the VLL model and a 300~GeV $\lpm$ in the type-III seesaw model. The error bars on the data points represent statistical uncertainties, and the shaded bands represent the systematic uncertainties on the background predictions.}
  \label{fig:SR-DeltaM}
\end{figure}

\begin{table}[h]
	\renewcommand{\arraystretch}{1.5}
	\centering
	\caption{Observed and expected number of events in the six signal regions, before and after the combined unbinned maximum-likelihood fit. The pre-fit uncertainties represent the total systematic uncertainties on the background estimates. The post-fit uncertainties are determined by the maximum-likelihood fit.}
	\sisetup{ table-number-alignment=center,
	separate-uncertainty=true,
	}
	\footnotesize
	\begin{tabular}{ | c ||
	S[separate-uncertainty,table-figures-uncertainty=1] |
	S[separate-uncertainty ,table-figures-uncertainty=1]|
	S[separate-uncertainty,table-figures-uncertainty=1] ||
	S[separate-uncertainty,table-figures-uncertainty=1] |
	S[separate-uncertainty,table-figures-uncertainty=1] |
	S[separate-uncertainty,table-figures-uncertainty=1] |
	}
		\hline
		&	\multicolumn{3}{c||}{$Z+e$}	&	\multicolumn{3}{c|}{$Z+\mu$}	\\
		\hline
		Process	&	{$4$l SR}	&	{$3l+jj$ SR}	&	{$3l$-only SR}	 &	{$4l$ SR}	&	{$3l+jj$ SR}	&	{$3l$-only SR}	\\
		\hline
		\hline
		& \multicolumn{6}{c|}{Before combined background-only fit} \\
		\hline

		$ZZ$	&	10.9 \pm 0.6	&	11.7 \pm 0.8	&	91 \pm 5	&	21.4 \pm 1.1	&	7.5 \pm 0.6	&	90 \pm 5	\\
		\hline
		$WZ$	&	0.08 \pm 0.01	&	35.3 \pm 3.1	&	337 \pm 28	 &  {\textemdash}	 & 	46 \pm 4	 & 	480 \pm 40	\\
		\hline
		$Z+\gamma$	&   {\textemdash}	 & 	2.3 \pm 0.8	 & 	35 \pm 11	 & 	  {\textemdash}	 & 	  {\textemdash}	 & 	{\textemdash}	\\
		\hline
		Reducible	 & 	  {\textemdash}	 & 	1.6 \pm 0.5	 & 	38 \pm 14	 & 	1.5 \pm 0.3	 & 	8.8 \pm 3.0	 & 	79 \pm 22	\\
		\hline
		$t\overline{t}+V, VVV^{(*)} $ &  1.2 \pm 0.2  &  7.8 \pm 1.7  &  2.3 \pm 0.4  &  1.5 \pm 0.2  &  9.5 \pm 2.1  &  3.3 \pm 0.5 \\
		\hline
		Total Background & 	12.2 \pm 0.7	 & 	59 \pm 4	 & 	504 \pm 34	 & 	24.4 \pm 1.2	 & 	72 \pm 6	 & 	650 \pm 50 	\\
		\hline
		\hline
		& \multicolumn{6}{c|}{After combined background-only fit} \\
		\hline
		$ZZ$ & 15 \pm 4 &   13.4\pm 2.3 &  107\pm 9 & 22 \pm 5 &  10.1 \pm 1.6 &  88 \pm 8 \\
		\hline
		$WZ$ & 0.08 \pm 0.03 &   39 \pm 6 &  393 \pm 28 & 0.02 \pm 0.02 &   56 \pm 9 &  460 \pm 40 \\
		\hline
		$Z + \gamma$ &   {\textemdash} &   2.2 \pm 0.8 &  34 \pm 11 &   {\textemdash} &    {\textemdash} &   {\textemdash} \\
		\hline
		Reducible &   {\textemdash}  &   1.8 \pm 1.2 &  37 \pm 13 & 1.8 \pm 0.9 &  10.2 \pm 2.8 &  92 \pm 24 \\
		\hline
		$t\overline{t}+V, VVV^{(*)}$  & 1.1\pm 0.3 &   7.5 \pm 1.7 &  2.5 \pm 0.6 & 1.5\pm 0.4 &  9.1 \pm 2.1 &  3.3 \pm 0.8 \\
		\hline
		Total Background & 16 \pm 4 & 64\pm7  & 574\pm 34 &  25\pm5  & 85\pm10 &  640\pm40 \\
		\hline
		\hline
		Data 	&	\multicolumn{1}{c|}{16}	&	\multicolumn{1}{c|}{64}	&	\multicolumn{1}{c||}{573}	&	\multicolumn{1}{c|}{25}	&	\multicolumn{1}{c|}{86}	&	\multicolumn{1}{c|}{651}	\\
		\hline
	\end{tabular} 

	\label{table:SR-event-counts}
\end{table}

\begin{figure}[htbp]
  \centering	
  \subfloat[ ] {
    \resizebox{0.48\textwidth}{!}{\includegraphics{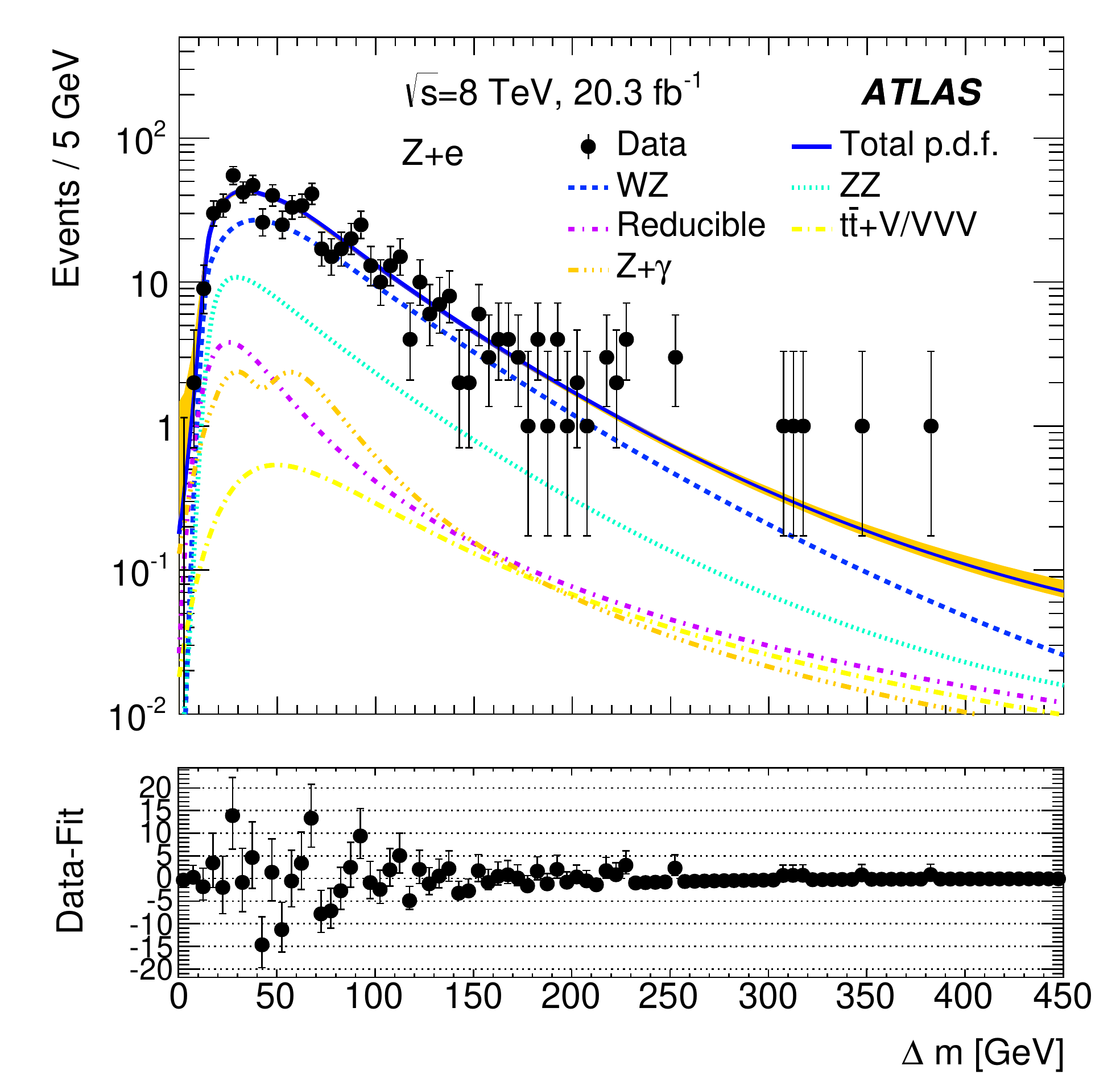}}
  }
  \subfloat[ ] {
    \resizebox{0.48\textwidth}{!}{\includegraphics{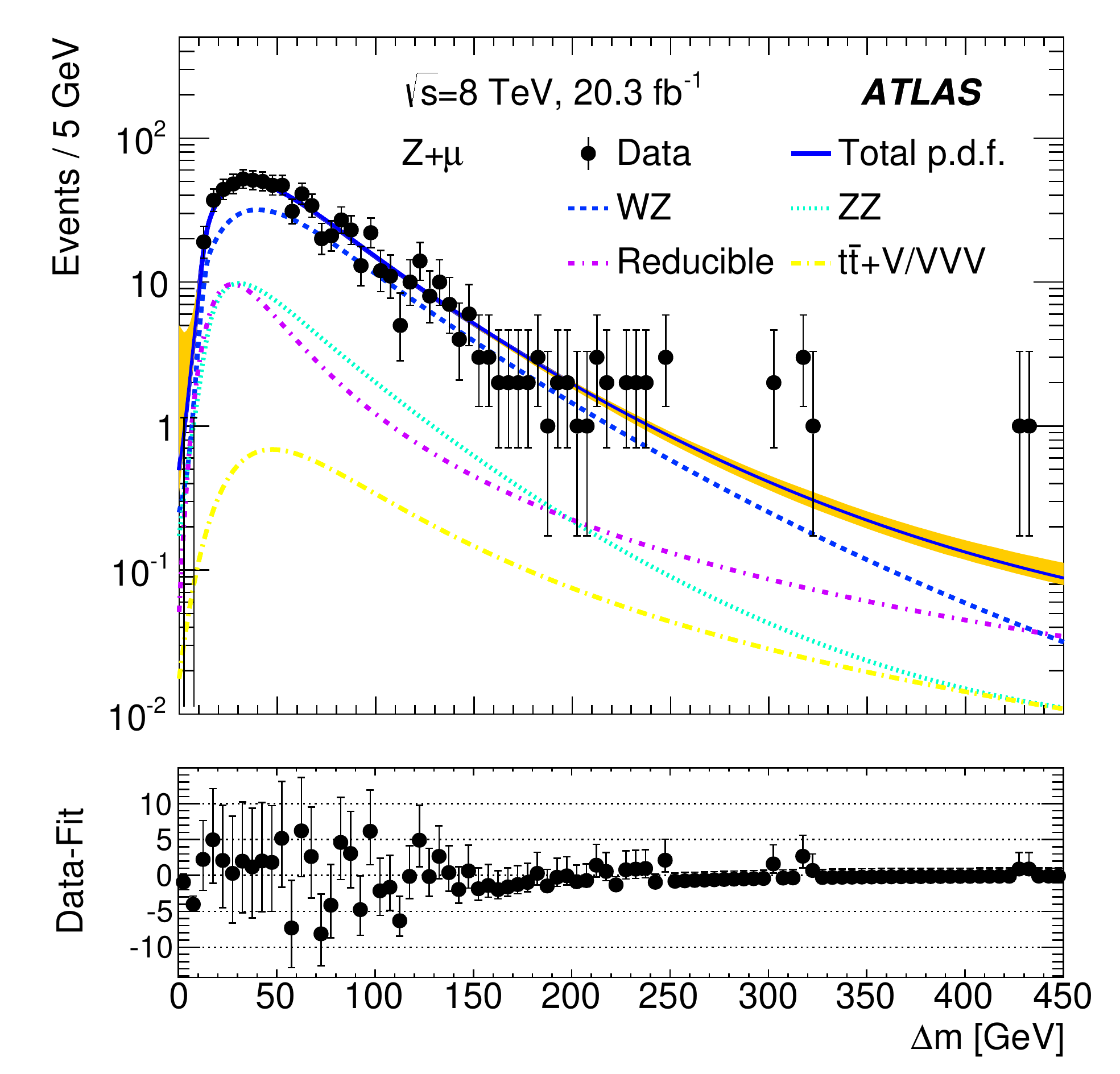}}
  } \\
  \caption{Projections onto the $\deltam$ variable of the background-only unbinned maximum-likelihood fits, shown superimposed on the data with the three categories in each flavour channel added together. The $Z+e$ flavour channel is shown in (a), and the $Z+\mu$ channel is shown in (b). The contributions of the separate background components to the total background-only fit are also shown. The error bars on the data points represent statistical uncertainties. Good agreement is observed between the background model and the data.}
  \label{fig:FitResult}
\end{figure}

Good agreement is seen between the pre-fit and post-fit normalizations for the $\fourl$ and $\threeljj$ categories in the $Z+\mu$ flavour channel. The largest change in normalization due to the fit is in the $\fourl$ category for the $Z+e$ flavour channel, where the fitted $ZZ$ normalization exceeds the prediction by $35\%$. The $WZ$ and $ZZ$ normalizations increase by roughly $15\%$ in the $\threeljj$ and $\threelo$ categories in the $Z+e$ flavour channel, and $30\%$ in the $\threeljj$ category in the $Z+\mu$ flavour channel. The projections of the fit results in the background-only hypothesis are shown in figure~\ref{fig:FitResult} for the combination of the three categories in each flavour channel.

The data are well described by the combined fit to the three categories in each flavour channel. 
The consistency of the data with the background-only hypothesis is evaluated by scanning the local $p_0$-value for the $\deltam$ distribution in $3 \GeV$ intervals for signal mass hypotheses in the range $100-400 \GeV$ for the VLL model, and $100-500 \GeV$ for the seesaw model, 
using the unbinned maximum-likelihood fit described in section~\ref{sec:fit-procedure} with the signal strength set to zero. The $p_0$-value, which corresponds to the probability to observe at least as many events as observed in the present measurement assuming the background-only hypothesis, is calculated using the frequentist hypothesis test based on the profile likelihood ratio test statistic and approximated with asymptotic formulae~\cite{asimov}. The minimum $p_0$-value is $p_0=0.02$ at a mass of $183~\GeV$ for the $Z+e$ flavour channel, and $p_0=0.05$ at a mass of $109~\GeV$ for the $Z+\mu$ flavour channel.

Since no significant excess above the background expectation is observed, the fit model is used to derive 95\% confidence level (CL) exclusion limits on the heavy lepton pair-production cross section, $\sigma$,
using the $CL_s$ method~\cite{CLs}. The limits are shown for the VLL model in figure~\ref{fig:VLLLimit}, and for the type-III seesaw model in figure~\ref{fig:T3SSLimit}, evaluated in the same $3 \GeV$ intervals as the $p_0$-values.
The VLL model is excluded for electron-only mixing in the heavy lepton mass ranges $129$--$144 \GeV$ and $163$--$176 \GeV$, with an expected exclusion in the range $109$--$152 \GeV$. The corresponding observed (expected) exclusion for the muon-only mixing scenario is $114$--$153 \GeV$ and $160$--$168 \GeV$ ($105$--$167 \GeV $).
The significantly higher production cross sections for the type-III seesaw model lead to an observed (expected) exclusion in the electron-only mixing scenario in the heavy lepton mass range $100$--$430 \GeV$  ($100$--$436 \GeV$). For the muon-only mixing scenario, the observed exclusion is in the ranges $100$--$401 \GeV$ and $419$--$468 \GeV$, while the expected exclusion is $100$--$419 \GeV$.

\begin{figure}[htbp]
  \centering	
  \subfloat[] {
    \resizebox{0.5\textwidth}{!}{\includegraphics{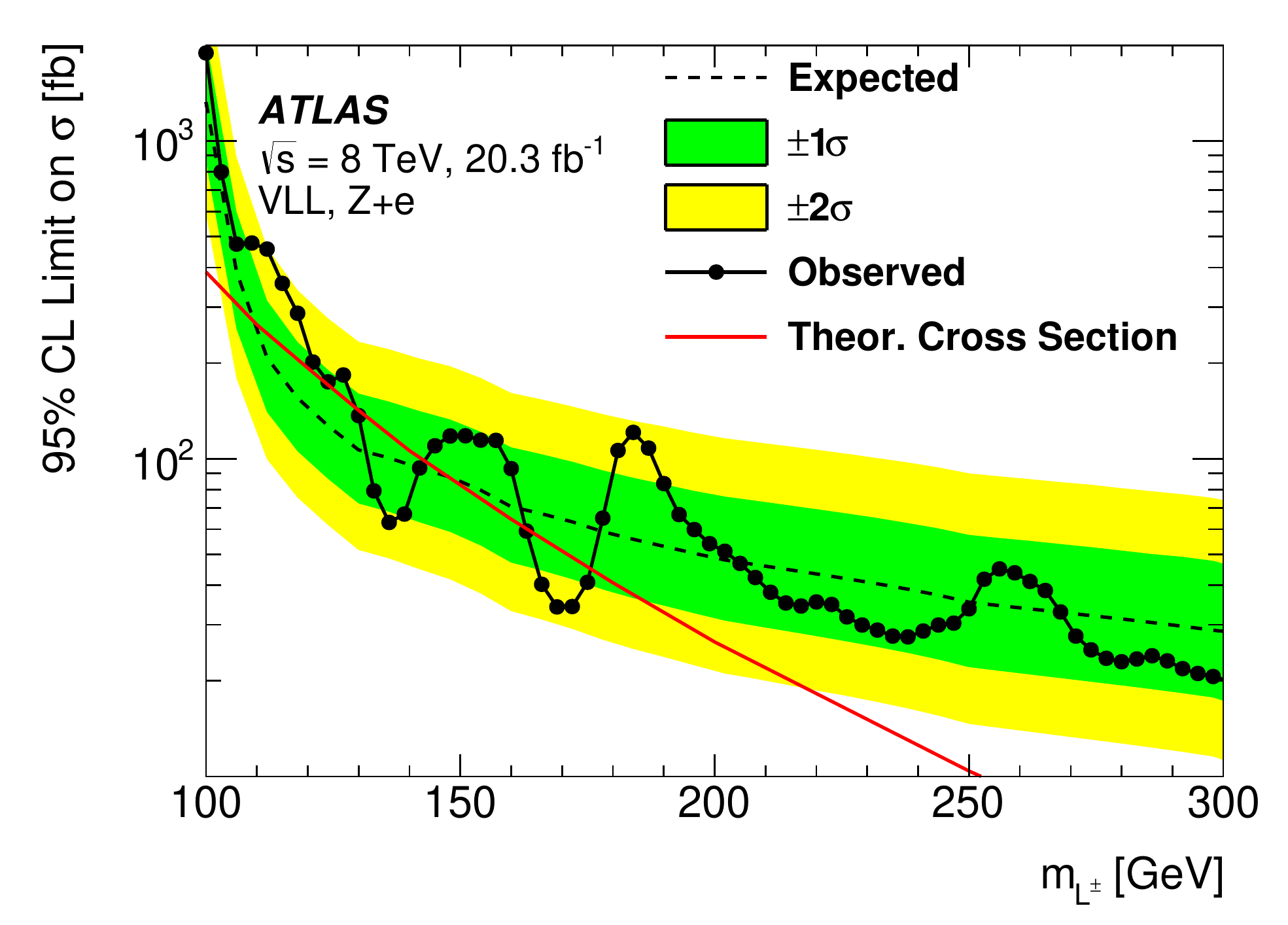}}
  }
  \subfloat[] {
    \resizebox{0.5\textwidth}{!}{\includegraphics{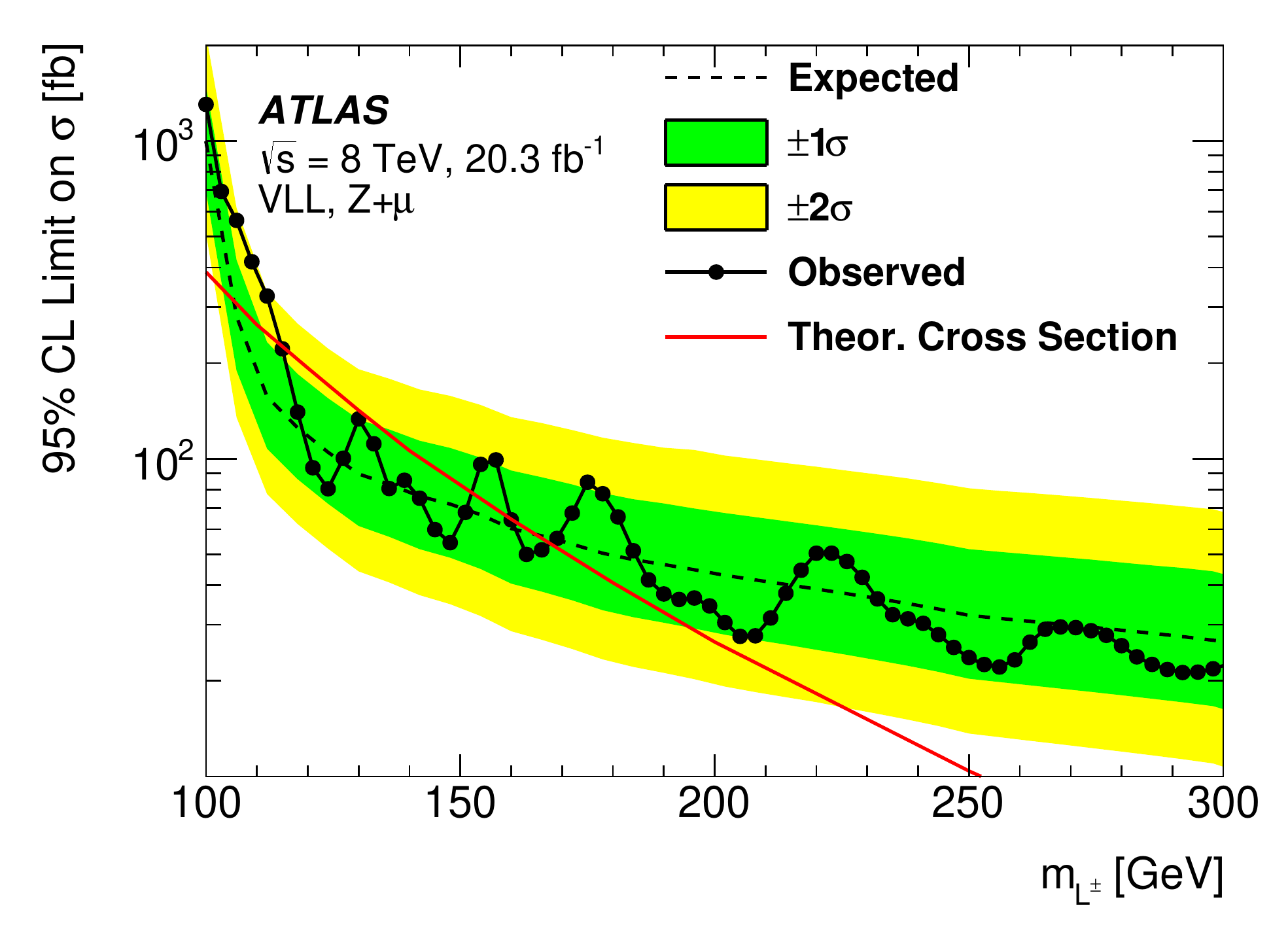}}
  } \\
  \caption{$95 \%$ CL upper limits on the vector-like lepton cross section. The left (right) plot shows the limits assuming $100\%$ branching fraction to $e/\nu_e$ ($\mu/\nu_{\mu}$). The solid line shows the observed limit. The dashed line shows the median expected limit for a background-only hypothesis, with green and yellow bands indicating the expected fluctuations at the $\pm1\sigma$ and $\pm 2 \sigma$ levels. The limit is evaluated in $3 \GeV$ intervals.}
  \label{fig:VLLLimit}
\end{figure}

\begin{figure}[htbp]
  \centering
  \subfloat[] {
     \resizebox{0.5\textwidth}{!}{\includegraphics{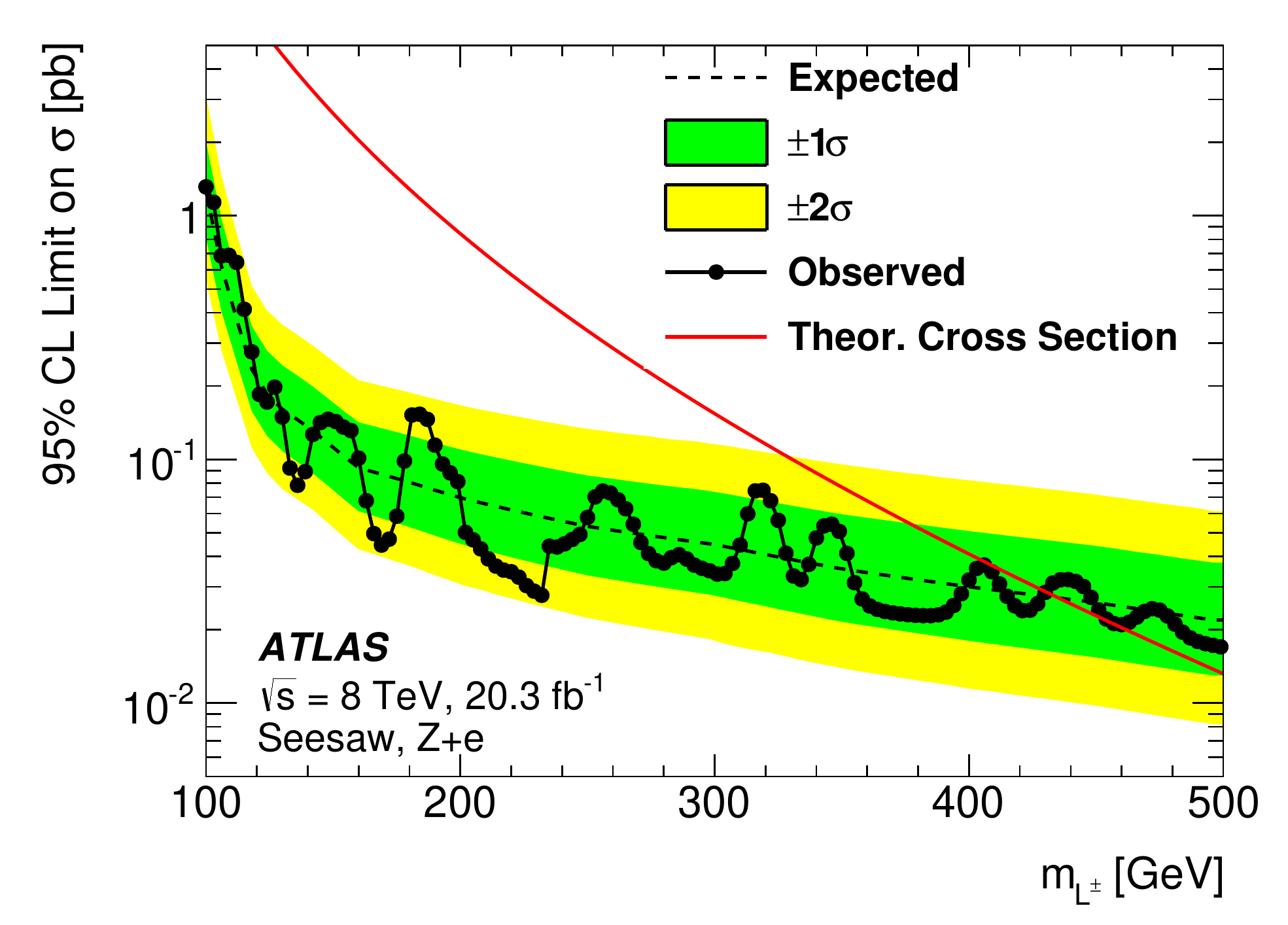}}
  }
  \subfloat[] {
     \resizebox{0.5\textwidth}{!}{\includegraphics{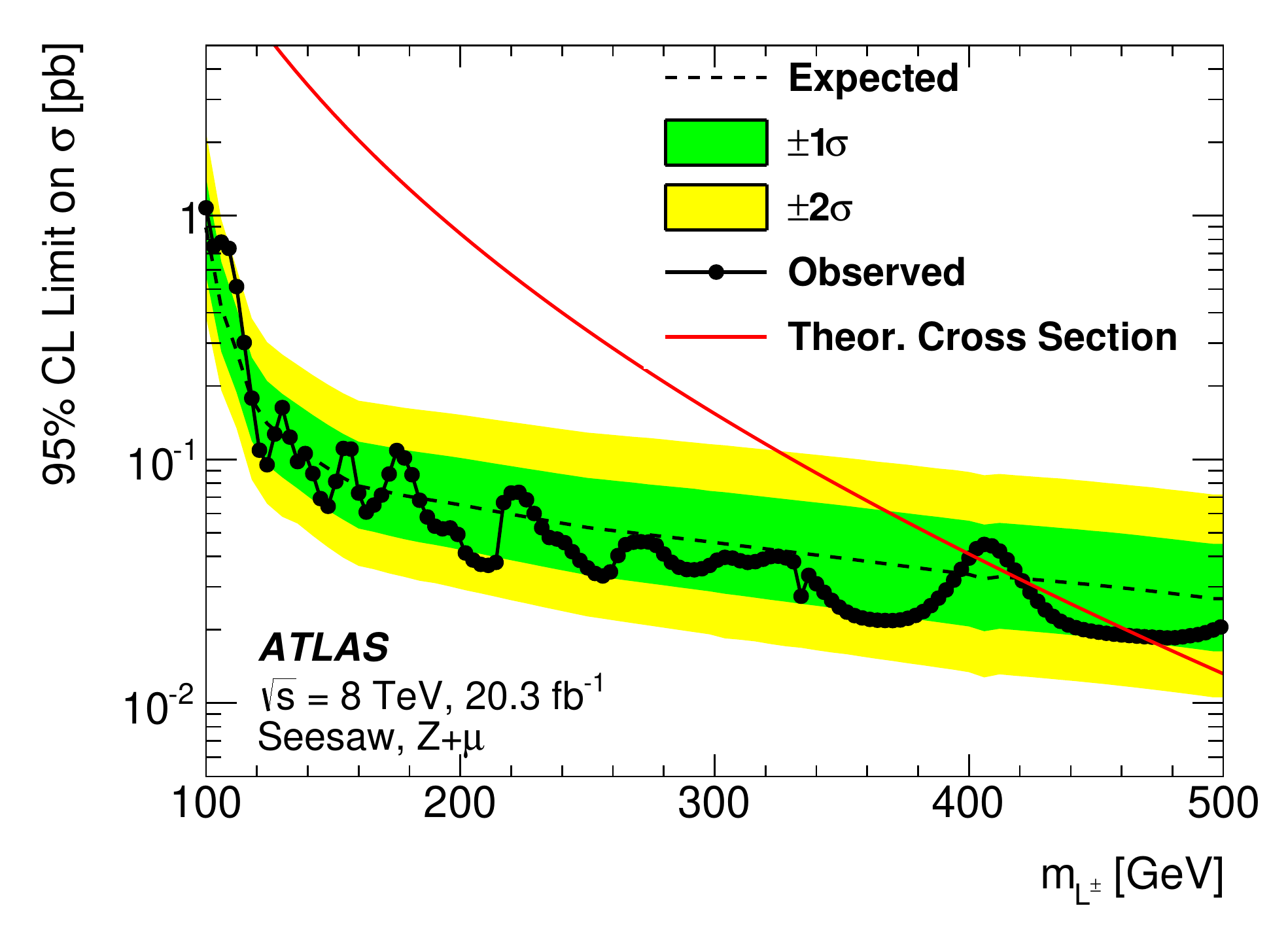}}
  }
  \caption{$95 \%$ CL upper limits on the type-III seesaw production cross section. The left (right) plot shows the limits assuming $100\%$ branching fraction to $e/\nu_e$ ($\mu/\nu_{\mu}$). The solid line shows the observed limit. The dashed line shows the median expected limit for a background-only hypothesis dataset, with green and yellow bands indicating the expected fluctuations at the $\pm1\sigma$ and $\pm 2 \sigma$ levels. The limit is evaluated in $3 \GeV$ intervals.}
  \label{fig:T3SSLimit}
\end{figure}

The constraints shown in figures~\ref{fig:VLLLimit} and \ref{fig:T3SSLimit} are relevant to the specific VLL and type-III seesaw models considered, and are not necessarily applicable to other scenarios predicting trilepton resonances with an intermediate $Z$ boson. A more model-independent observable is the \emph{visible cross section}, $\sigmavis$, defined as the number of observed events with $Z+\ell$-induced trilepton resonances for a given resonance mass divided by the integrated luminosity of the data sample, 20.3~\ifb. The 95\% CL upper limits on $\sigmavis$, denoted $\sigma_{95}^{\mathrm{vis}}$, are derived from a fit to each flavour channel with $f_V=1$, i.e. using only the peak component of the signal. The results for the two flavour channels, derived from the inclusive event selection without dividing the events into the three categories, are shown in figure~\ref{fig:N95}.

The limits on $\sigmavis$ can be used to test specific models after taking into account the model's acceptance with respect to a fiducial volume, $\mathcal{A}$, and reconstruction and selection efficiency of events within the fiducial volume, $\epsilon_{\mathrm{fid}}$. The 95\% CL upper limit on the cross section for the model is given by:

\begin{equation}\label{eqn:sigma95}
	\sigma_{95} = \frac{\sigma_{95}^{\mathrm{vis}}}{\mathcal{A}\times \epsilon_{\mathrm{fid}}}.
\end{equation}

The acceptance $\mathcal{A}$ is defined as the probability for generated signal events to lie within a fiducial volume defined by the kinematics of the generated leptons. The leptons are considered at \emph{particle level}, i.e. after parton shower and hadronization and with lifetimes longer than $10^{-11}$~s, and are \emph{dressed}, including the contributions from radiated photons within a cone of $\Delta R=0.1$. The fiducial volume requires that events contain an $\lpm$ decaying to a prompt electron or muon and a $Z$ boson that then decays to electrons or muons. The three leptons from the $\lpm$ decay are required to have $\pt>15~\GeV$ and lie within $|\eta|<2.5$, with at least one lepton satisfying $\pt>26~\GeV$. Two of the leptons must form a same-flavour opposite-sign pair with a mass within $10 \GeV$ of $m_Z$, and the $Z$ boson and the off-$Z$ lepton must be separated by $\Delta R < 3$.  The events are divided into flavour channels according to the flavour of the off-$Z$ lepton. For the VLL and type-III seesaw models used in this analysis, the acceptance of events containing an $\lpm\rightarrow Z(\ell\ell)\ell$ decay to fall within the fiducial volume is in the range $60\%$--$65\%$ for most of the mass range, decreasing at higher masses due to the cut on the $\Delta R$ between the $Z$ boson and the off-$Z$ lepton. The acceptance decreases at low masses due to the lepton $\pt$ requirement, reaching $30\%$--$35\%$ at $m_{\lpm}=100 \GeV$.

For type-III seesaw and VLL events within the fiducial volume, $\epsilon_{\mathrm{fid}}$ ranges from 20\% to 49\% if the other heavy lepton decays to a neutrino and a $W$, $Z$, or $H$ boson~\footnote{Note that the quoted efficiencies are dependent on the modelling of the polarization of the $Z$ bosons, due to the requirements imposed on lepton isolation and separation.}. If the other heavy lepton decays to an electron or a muon, the efficiency is 10\%--20\% lower, due to the increased probability of incorrectly selecting the off-$Z$ lepton. The event selection efficiencies for the type-III seesaw model in scenarios where the second heavy lepton decays to a $W$ boson are shown in figure~\ref{fig:fiducial-efficiencies-inclusive} as a function of $m_{\lpm}$; the efficiencies for scenarios where the second heavy lepton decays to a $Z$ or $H$ boson and for the VLL model are consistent with these efficiencies within the statistical uncertainties.

\begin{figure}[htbp]
  \subfloat[] {
  	\includegraphics[width=0.48\columnwidth]{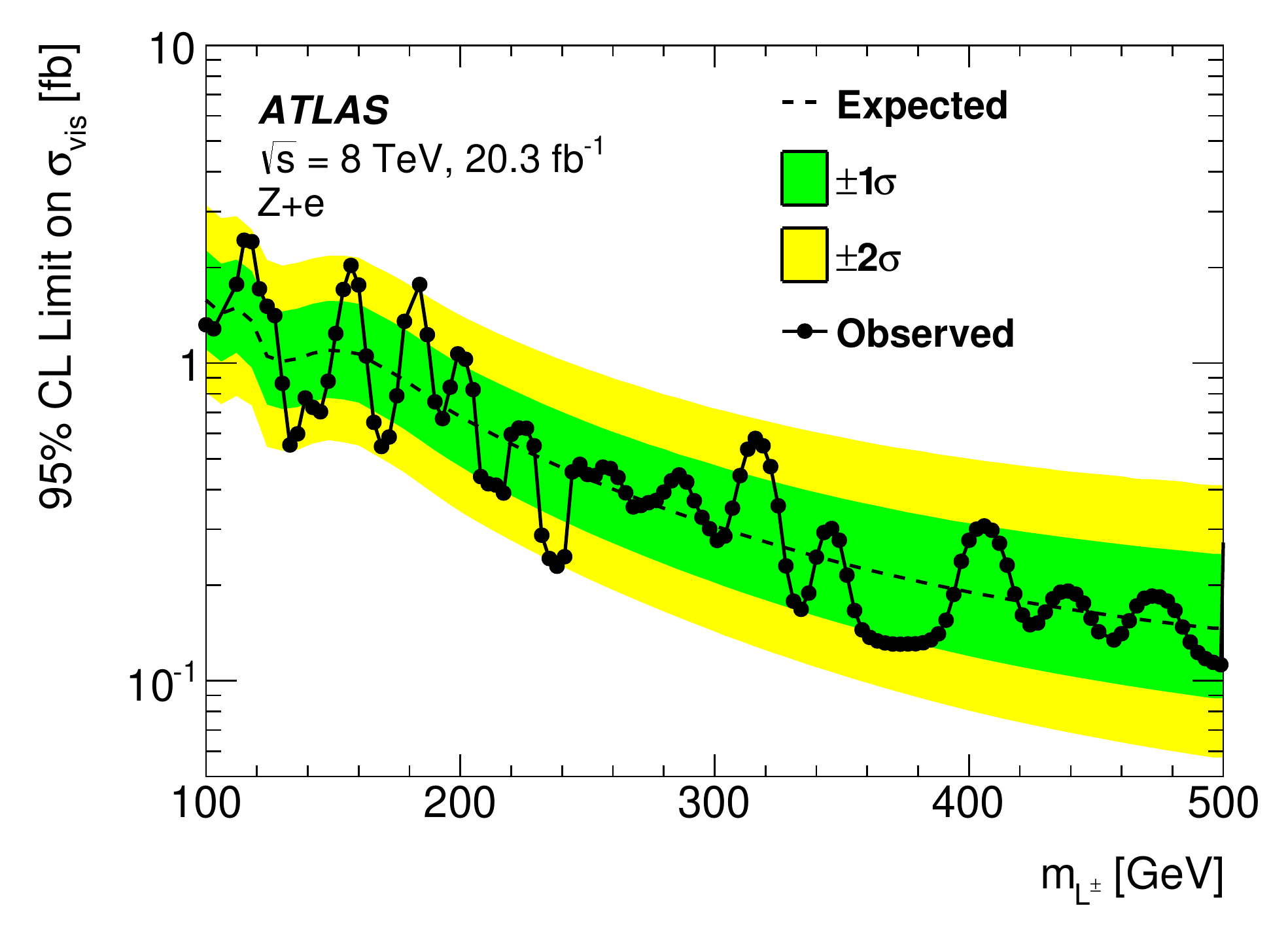}
  }
  \subfloat[] {
  	\includegraphics[width=0.48\columnwidth]{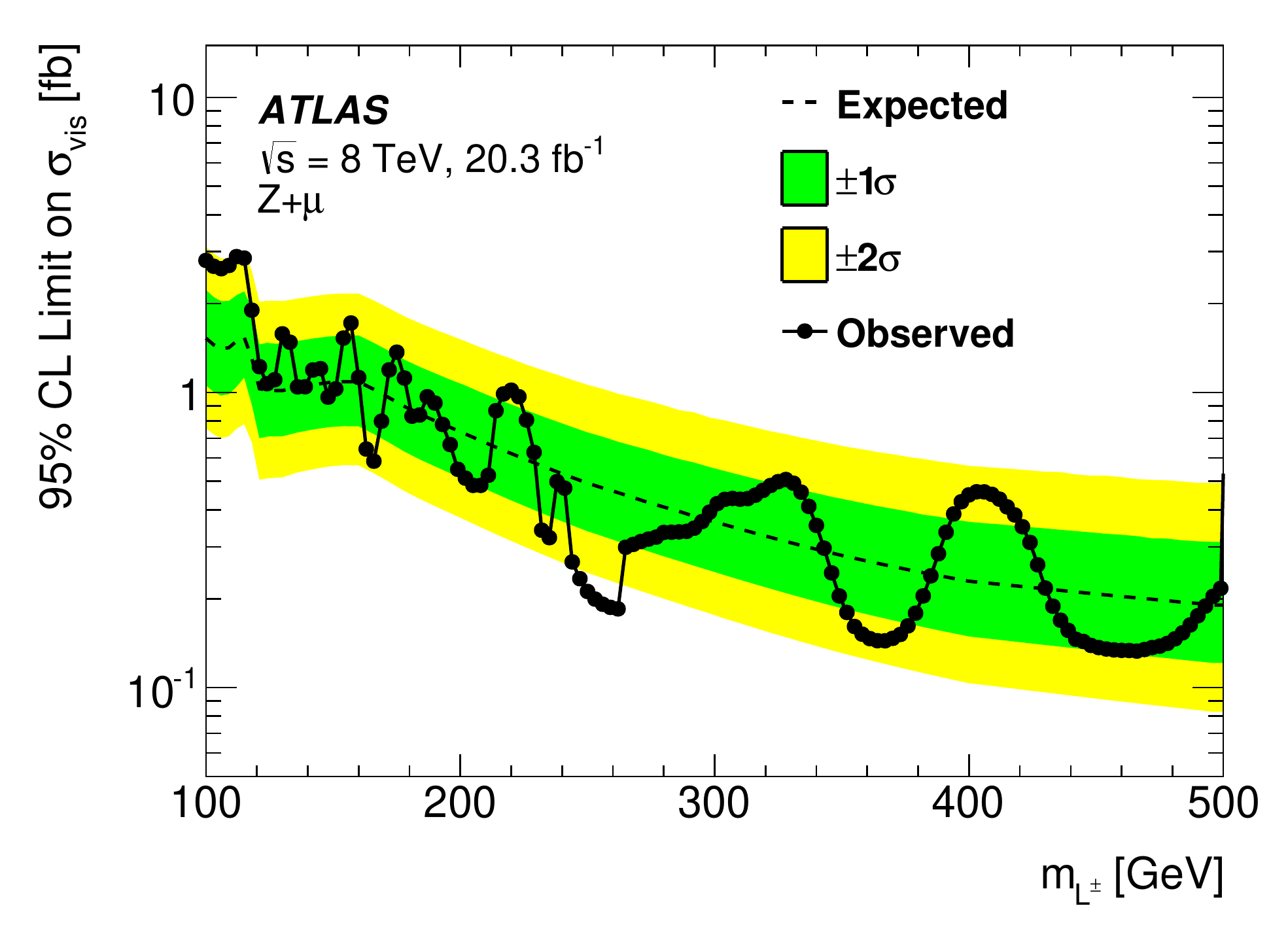}
  }
  \caption{Upper limits at 95\% CL on $\sigmavis$ for the $Z+e$ (left) and $Z+\mu$ (right) flavour channels, derived without dividing events into the three categories. The limits are evaluated in $3 \GeV$ intervals.}
  \label{fig:N95}
\end{figure}

\begin{figure}[htbp]
  \subfloat[] {
  	\includegraphics[width=0.48\columnwidth]{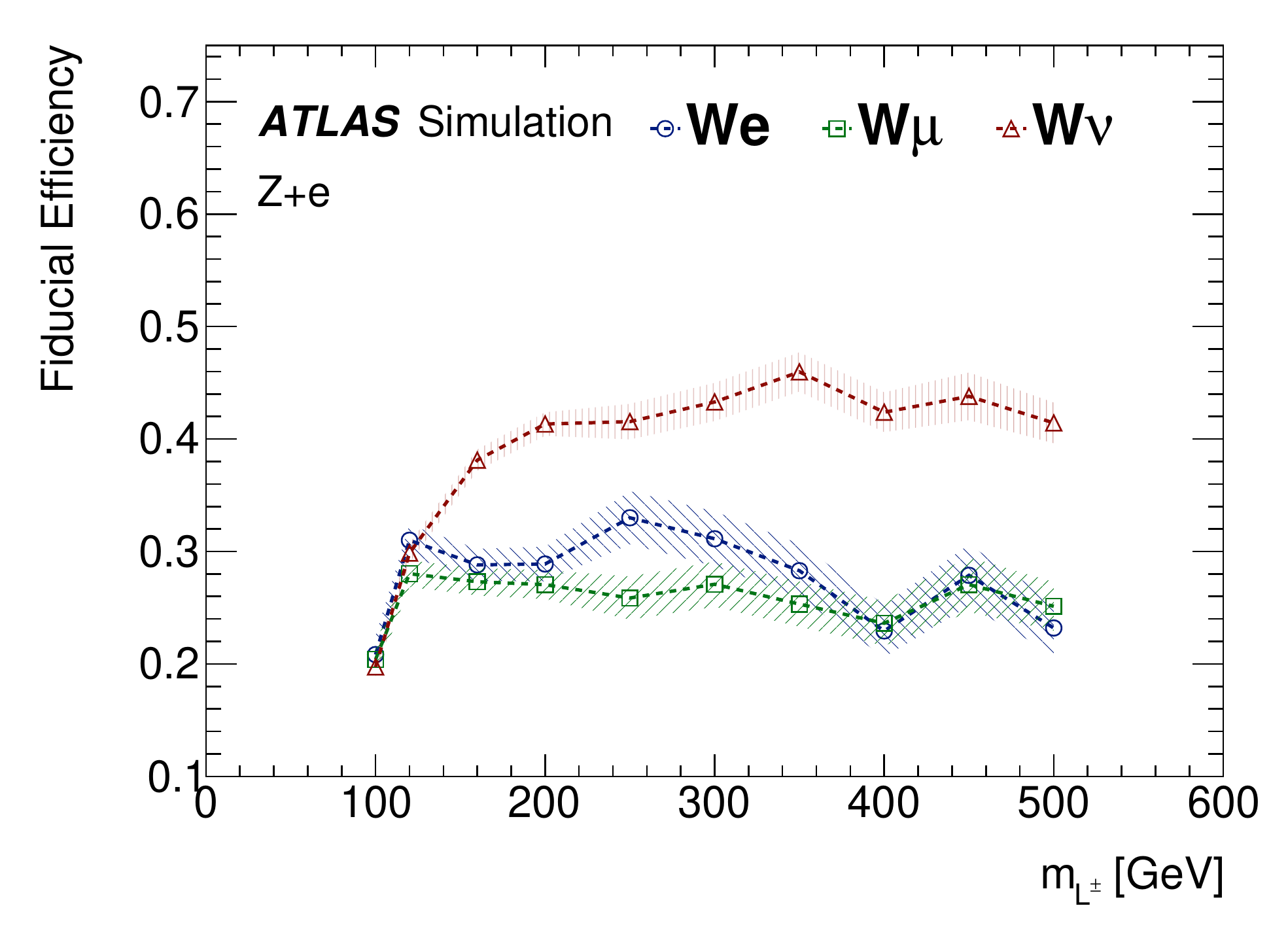}
  }
  \subfloat[] {
  	\includegraphics[width=0.48\columnwidth]{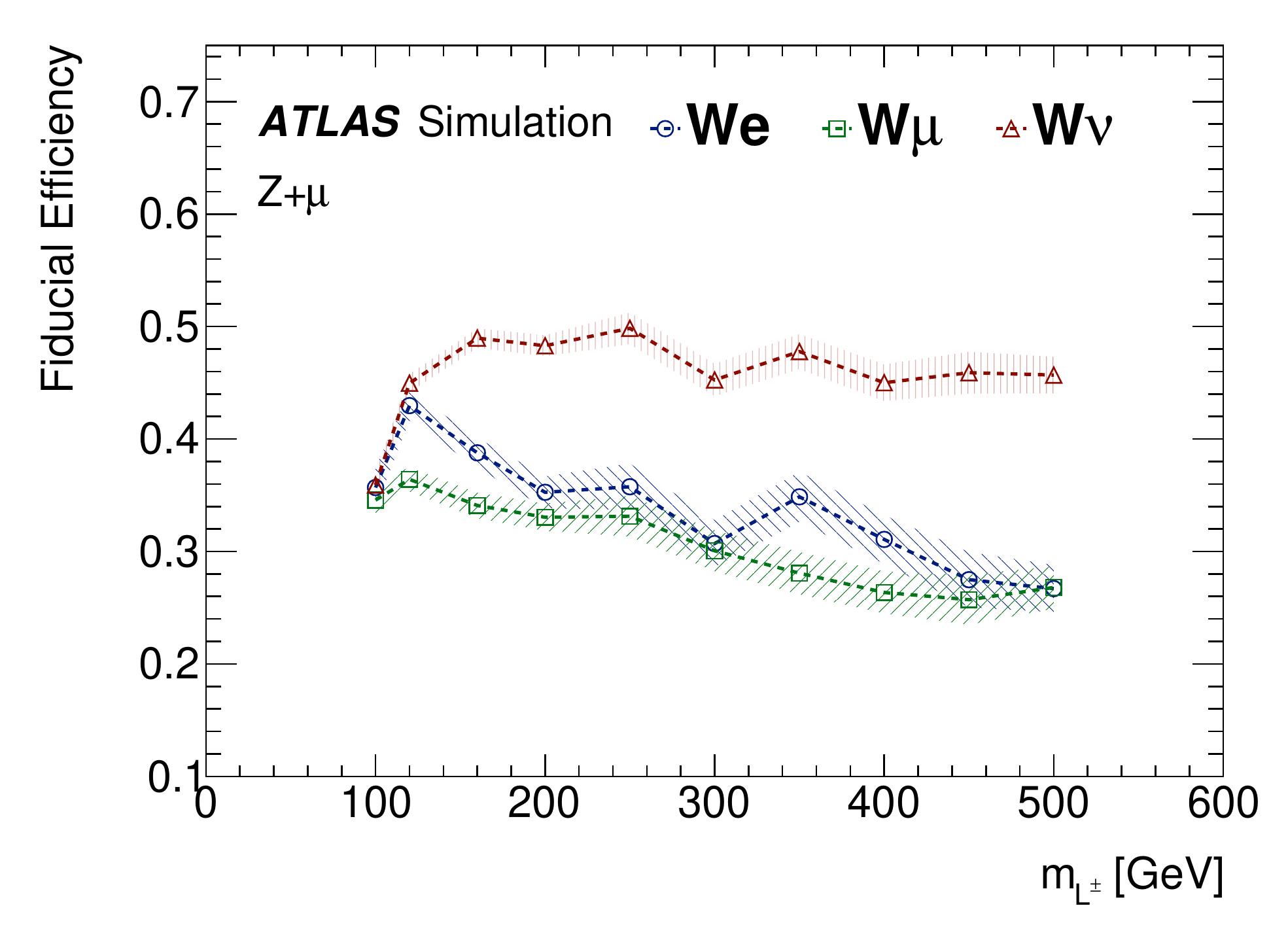}
  }
  \caption{Efficiencies for reconstructing and correctly identifying the $\lpm\rightarrow Z(\ell\ell)\ell^{\pm}$ decay in events within the fiducial volume for the type-III seesaw model. The left (right) plot shows the efficiencies for events containing a $\lpm\rightarrow Z(\ell\ell)e$ ($\lpm\rightarrow Z(\ell\ell)\mu$) decay. The decay of the second heavy lepton is specified in the legend. The shaded bands show the statistical uncertainty.}
  \label{fig:fiducial-efficiencies-inclusive}
\end{figure}

\FloatBarrier

\section{Conclusion}
\label{sec:conclusion}

A search for trilepton resonances decaying to a $Z$ boson and an electron or a muon has been presented. The search is based on $pp$ collision data taken at $\sqrt{s}=8$~TeV corresponding to an integrated luminosity of 20.3~fb$^{-1}$, collected by the ATLAS experiment at the CERN Large Hadron Collider. Events are selected requiring at least three electrons or muons with high transverse momentum, with two of the leptons consistent with the decay of a $Z$ boson. The events are categorized based on the presence or absence of additional leptons or dijet pairs in the event consistent with the decay products of a second heavy lepton, and separated into channels based on the flavour of the lepton associated with the $Z$ boson to form a heavy lepton decay candidate. Using the difference between the trilepton and the $Z$ boson candidate masses, a search for a narrow resonance is performed in each of these categories using a maximum-likelihood fit of parameterized signal and background shapes to the data.  No significant excess above Standard Model predictions is observed, and 95\% CL limits on the production of trilepton resonances beyond the Standard  Model are derived.  The results are interpreted in the context of two models of new heavy leptons decaying to three charged leptons.  In the vector-like lepton model, new heavy charged leptons are excluded in the mass range $129$--$176 \GeV$ ($114$--$168 \GeV$) for electron-only (muon-only) mixing, except for the interval $144$--$163 \GeV$ ($153$--$160 \GeV$). In the type-III seesaw model, the corresponding exclusion is in the mass range $100$--$430 \GeV$ ($100$--$468 \GeV$) for electron-only (muon-only) mixing, except for the interval $401$--$419 \GeV$ in the muon case. Limits are also set on the visible cross section of trilepton resonance productions, and fiducial efficiencies are derived to facilitate model testing.

\section*{Acknowledgements}

We thank CERN for the very successful operation of the LHC, as well as the
support staff from our institutions without whom ATLAS could not be
operated efficiently.

We acknowledge the support of ANPCyT, Argentina; YerPhI, Armenia; ARC,
Australia; BMWFW and FWF, Austria; ANAS, Azerbaijan; SSTC, Belarus; CNPq and FAPESP,
Brazil; NSERC, NRC and CFI, Canada; CERN; CONICYT, Chile; CAS, MOST and NSFC,
China; COLCIENCIAS, Colombia; MSMT CR, MPO CR and VSC CR, Czech Republic;
DNRF, DNSRC and Lundbeck Foundation, Denmark; EPLANET, ERC and NSRF, European Union;
IN2P3-CNRS, CEA-DSM/IRFU, France; GNSF, Georgia; BMBF, DFG, HGF, MPG and AvH
Foundation, Germany; GSRT and NSRF, Greece; RGC, Hong Kong SAR, China; ISF, MINERVA, GIF, I-CORE and Benoziyo Center, Israel; INFN, Italy; MEXT and JSPS, Japan; CNRST, Morocco; FOM and NWO, Netherlands; BRF and RCN, Norway; MNiSW and NCN, Poland; GRICES and FCT, Portugal; MNE/IFA, Romania; MES of Russia and NRC KI, Russian Federation; JINR; MSTD,
Serbia; MSSR, Slovakia; ARRS and MIZ\v{S}, Slovenia; DST/NRF, South Africa;
MINECO, Spain; SRC and Wallenberg Foundation, Sweden; SER, SNSF and Cantons of
Bern and Geneva, Switzerland; NSC, Taiwan; TAEK, Turkey; STFC, the Royal
Society and Leverhulme Trust, United Kingdom; DOE and NSF, United States of
America.

The crucial computing support from all WLCG partners is acknowledged
gratefully, in particular from CERN and the ATLAS Tier-1 facilities at
TRIUMF (Canada), NDGF (Denmark, Norway, Sweden), CC-IN2P3 (France),
KIT/GridKA (Germany), INFN-CNAF (Italy), NL-T1 (Netherlands), PIC (Spain),
ASGC (Taiwan), RAL (UK) and BNL (USA) and in the Tier-2 facilities
worldwide.

\printbibliography

\newpage
\begin{flushleft}
{\Large The ATLAS Collaboration}

\bigskip

G.~Aad$^{\rm 85}$,
B.~Abbott$^{\rm 113}$,
J.~Abdallah$^{\rm 151}$,
O.~Abdinov$^{\rm 11}$,
R.~Aben$^{\rm 107}$,
M.~Abolins$^{\rm 90}$,
O.S.~AbouZeid$^{\rm 158}$,
H.~Abramowicz$^{\rm 153}$,
H.~Abreu$^{\rm 152}$,
R.~Abreu$^{\rm 30}$
% ...
\bigskip
\\
% ...
$^{11}$ Institute of Physics, Azerbaijan Academy of Sciences, Baku, Azerbaijan\\
% ...
$^{30}$ CERN, Geneva, Switzerland\\
% ...
$^{85}$ CPPM, Aix-Marseille Universit{\'e} and CNRS/IN2P3, Marseille, France\\
% ...
$^{90}$ Department of Physics and Astronomy, Michigan State University, East Lansing MI, United States of America\\
% ...
$^{107}$ Nikhef National Institute for Subatomic Physics and University of Amsterdam, Amsterdam, Netherlands\\
% ...
$^{113}$ Homer L. Dodge Department of Physics and Astronomy, University of Oklahoma, Norman OK, United States of America\\
% ...
$^{151}$ Institute of Physics, Academia Sinica, Taipei, Taiwan\\
$^{152}$ Department of Physics, Technion: Israel Institute of Technology, Haifa, Israel\\
$^{153}$ Raymond and Beverly Sackler School of Physics and Astronomy, Tel Aviv University, Tel Aviv, Israel\\
% ...
$^{158}$ Department of Physics, University of Toronto, Toronto ON, Canada\\
% ...
\end{flushleft}


\end{document}